\documentclass[pra,reprint,superscriptaddress,showpacs]{revtex4-1}
\usepackage[T1]{fontenc}
\usepackage[latin9]{inputenc}
\usepackage{amsmath}
\usepackage{amssymb}
\usepackage{wasysym}
\usepackage{graphicx}
\usepackage[unicode=true,pdfusetitle,
 bookmarks=true,bookmarksnumbered=false,bookmarksopen=false,
 breaklinks=false,pdfborder={0 0 0},backref=false,colorlinks=false]
 {hyperref}

\begin{document}

\title{Parity measurement of remote qubits using dispersive coupling and
photodetection}

\author{J. Govenius}

\email{joonas.govenius@aalto.fi}

\affiliation{QCD Labs, COMP Centre of Excellence, Department of Applied Physics,
Aalto University, P.O. Box 13500, FIN-00076 Aalto, Finland}

\author{Y. Matsuzaki}

\affiliation{NTT Basic Research Laboratories, NTT Corporation, Kanagawa 243-0198,
Japan}

\author{I. G. Savenko}

\affiliation{QCD Labs, COMP Centre of Excellence, Department of Applied Physics,
Aalto University, P.O. Box 13500, FIN-00076 Aalto, Finland}

\author{M. Möttönen}

\affiliation{QCD Labs, COMP Centre of Excellence, Department of Applied Physics,
Aalto University, P.O. Box 13500, FIN-00076 Aalto, Finland}

\begin{abstract}
Parity measurement is a key step in many entanglement generation and
quantum error correction schemes. We propose a protocol for non-destructive
parity measurement of two remote qubits, i.e., macroscopically separated
qubits with no direct interaction. The qubits are instead dispersively
coupled to separate resonators that radiate to shared photodetectors.
The scheme is deterministic in the sense that there is no fundamental
upper bound on the success probability. In contrast to previous proposals,
our protocol addresses the scenario where number resolving photodetectors
are available but the qubit--resonator coupling is time-independent
and only dispersive.
\end{abstract}


\maketitle

\section{Introduction}

\global\long\def\bra#1{\left\langle #1\right|}
\global\long\def\ket#1{\left|#1\right\rangle }
\global\long\def\bket#1{\big|#1\big\rangle}
\global\long\def\Bket#1{\Big|#1\Big\rangle}
\global\long\def\braket#1#2{\left\langle #1\left|#2\right.\right\rangle }
\global\long\def\bbraket#1#2{\big\langle#1\big|#2\big\rangle}
\global\long\def\expect#1#2{\left\langle #1\left|#2\right|#1\right\rangle }
\global\long\def\bexpect#1#2{\big\langle#1\big|#2\big|#1\big\rangle}
\global\long\def\matrixel#1#2#3{\left\langle #1\left|#2\right|#3\right\rangle }
\global\long\def\matrixelsub#1#2#3#4{_{#4}\left\langle #1\left|#2\right|#3\right\rangle _{#4}}

 One of the main challenges in quantum computing is that generally controlled
qubit--qubit interactions should be strong while uncontrolled interactions
should be negligible. Conceptually, the most straightforward approach
to solving this issue is to physically remove uncontrolled degrees
of freedom within a distance comparable to the qubit spacing, e.g.,
by trapping ions in ultra-high vacuum~\cite{Cirac1995Quantum}. An
alternative approach is to entangle the qubits with photons that act
as flying ancilla qubits. This approach allows placing the stationary
qubits in remote locations because the photonic ancilla qubits generally
interact weakly with the environment. In order to entangle the stationary
qubits, the ancilla qubits only need to interfere optically with each
other at detector inputs \cite{Cabrillo1999Creation,Bose1999Proposal}.
Post-selection or local qubit operations conditioned on the detector
outputs can then entangle the stationary qubits \cite{Cabrillo1999Creation,Bose1999Proposal,Duan2003Efficient,Browne2003Robust,Barrett2005Efficient,Lim2005RepeatUntilSuccess,Ladd2006Hybrid,Busch2008Entangling,Matsuzaki2011Entangling,Azuma2012Quantum,Bernien2013Heralded,Bruschi2014Repeatuntilsuccess}.
Such entanglement generation
is at the heart of quantum repeaters and cluster-state models of quantum
computing \cite{Briegel1998Quantum,Kok2007Linear}.

More generally, the photon-mediated approach allows performing a remote
non-destructive parity measurement (RNPM) \cite{Azuma2012Quantum}
described by the measurement operators $\hat{\Pi}_{1}=\ket{gg}\bra{gg}+\ket{ee}\bra{ee}$
and $\hat{\Pi}_{-1}=\ket{ge}\bra{ge}+\ket{eg}\bra{eg}$ for the even
and odd parity outcomes, respectively. Here $\ket g$ and $\ket e$
are orthogonal single-qubit states that define the computational basis.
Parity measurement is more general than entanglement generation in the sense that
RNPM can generate entanglement without ancilla qubits,
but protocols for preparing entangled states do not necessarily allow measuring parity
without the use of ancilla qubits.
Specifically,
Bell
pairs are generated from an initial product state $\left(\ket g+\ket e\right)\otimes\left(\ket g+\ket e\right)/2$
by measuring its parity, which provides just the right amount of information
to produce maximally entangled pairs.
However, we envision that
the most useful application for ancilla-free
RNPM may be in quantum error correction, where multi-qubit
parity measurements are central to a wide variety of stabilizer codes
\cite{Terhal2015Quantum}. Remote parity measurement in particular
may prove useful in circumventing limitations
that are inherent to codes
that only use geometrically
local parity checks \cite{Bravyi2010Tradeoffs}.

We note that the main challenge in non-destructive parity measurement is
that the protocol must preserve
coherence of arbitrary superpositions within the parity subspaces.
In contrast, entanglement generation protocols drive the system into a known state
and may assume a fixed initial state.
In this sense RNPM is similarly challenging as applying a CPHASE gate
to remote qubits \cite{Lim2005RepeatUntilSuccess},
although the two operations are not interchangeable without ancilla qubits.

We propose a protocol for RNPM that can be implemented in circuit
quantum electrodynamics (cQED) \cite{Blais04,Wallraff2004Strong,Devoret2013Superconducting,Kelly2015State}
using standard and minimalistic resources, with the exception of number
resolving photodetectors with high temporal resolution. Such detectors
for itinerant microwave photons have not been realized to date but
are under active theoretical and experimental development \cite{Romero2009Microwave,Chen2011Microwave,Peropadre2011Approaching,Fan2013Breakdown,Hoi2013Giant,Sathyamoorthy2014Quantum,Fan2014Nonabsorbing,Govenius2014Microwave,Gasparinetti2015Fast,Koshino2015Theory}.
In addition to the detectors, the requirements for our protocol are
the following: a beam splitter, two qubit--resonator systems with
dispersive coupling, coherent drive pulses applied to the resonators,
and single-qubit phase gates conditioned on the recorded arrival times
of the photons.

We emphasize that our protocol places few restrictions on the systems used as qubits.
Specifically, we do not require $\Lambda$\nobreakdash-type
internal level structure to entangle the qubit state and a Fock state
of the resonator, which is the typical starting point of most proposals
for optical experiments \cite{Cabrillo1999Creation,Bose1999Proposal,Duan2003Efficient,Browne2003Robust,Barrett2005Efficient,Lim2005RepeatUntilSuccess,Busch2008Entangling,Bernien2013Heralded,Bruschi2014Repeatuntilsuccess}.
Instead, our proposal works with a time-independent dispersive shift
$\chi$ and resonator decay rate $\kappa$,
both of which are standard features of cQED setups.
The time-independent and dispersive
qubit--resonator coupling gradually entangles the qubit state
with coherent states of the resonator
and continues to play an important role throughout the protocol.
This is in contrast to the generation of single photons on demand in Ref.~\cite{Lim2005RepeatUntilSuccess}, where the qubit state is entangled with Fock states
effectively instantaneously compared to other time scales.
However, if on-off modulation
of $\chi$ is possible, we can use it to speed up the parity measurement
by turning off the interaction at a specific time $t_{\textrm{off}}$
that maximizes the qubit--resonator entanglement. If the strong dispersive
limit \cite{Schuster2007Resolving} $\chi\gg\kappa$ is also reached,
the interaction becomes effectively instantaneous compared to other
time scales ($t_{\textrm{off}}\ll\kappa^{-1}$). In this particular
scenario, the protocol after $t_{\textrm{off}}$ reproduces a specific
case of the protocol proposed by Azuma et al. in Ref.~\cite{Azuma2012Quantum}.

Kerckhoff et al. proposed another parity measurement protocol that
relies on homodyne detection and sequential reflection of a probe
signal from two resonators coupled to three-level atoms \cite{Kerckhoff2009Physical}.
Roch et al. experimentally demonstrated the generation of odd-parity
Bell states in cQED using a similar sequential setup, but with dispersive
qubit--resonator coupling and three distinct outcomes \cite{Roch2014Observation}.
The disadvantage of the latter scheme is that it distinguishes the even
parity states from each other and therefore reveals too much to function
as RNPM. Our protocol on the other hand measures both parities non-destructively
in a non-sequential setup. This is possible because the photodetectors
effectively erase the phase information that would allow distinguishing
qubit states of the same parity. Instead, the photodetectors reveal
the stochastic relative phase acquired by the states, which is the
fundamental backaction of dispersive measurements \cite{Schuster2005Ac,Gambetta2006Qubitphoton,Gambetta2008Quantum}.
Conditioning phase gates on the photodetector outputs therefore allows
undoing the measurement-induced dephasing within the parity subspaces,
much like in the extensively studied case of a joint measurement of
two qubits in a single resonator \cite{Lalumiere2010Tunable,Tornberg2010Highfidelity,Riste2013Deterministic}.

The remainder of this article is organized as follows. Section~\ref{sec:1q}
reviews measurement-induced dephasing and discusses its reversal
in a single-qubit scenario. Section~\ref{sec:parity} introduces the
RNPM protocol and demonstrates its validity under ideal conditions.
It also suggests an alternative variation of the protocol for the
strong dispersive limit and briefly discusses some practical hurdles
to implementing the protocol. Section~\ref{sec:conclusion} concludes
the article.

\section{Reversing measurement-induced dephasing of a single qubit\label{sec:1q}}

\begin{figure}
\includegraphics{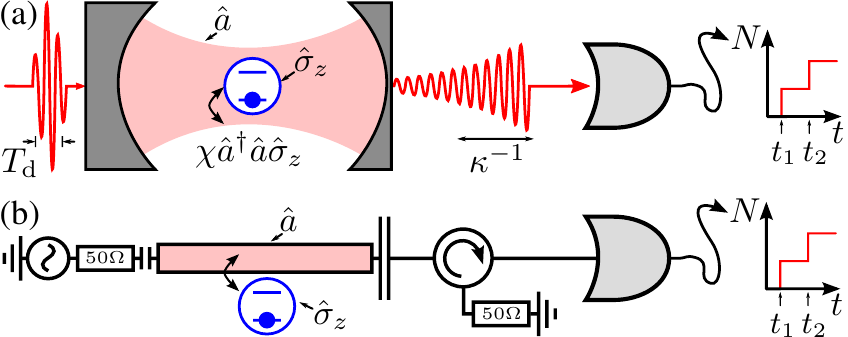}
\caption{(Color online) (a) Schematic diagram of a qubit dispersively coupled to
a resonator that radiates into a photodetector. At $t=0$, a short
($T_{\textrm{d}}\ll\kappa^{-1},\chi^{-1}$) external pulse drives
the resonator mode ($\hat{a}$) from vacuum to a coherent state $\ket{\alpha}$.
The resonator then evolves into a superposition of two coherent states
at a rate $\chi$ due to a dispersive interaction that entangles the
qubit ($\hat{\sigma}_{z}$) and the phase of the coherent state. At
a rate $\kappa$, the resonator decays through a single port to an
output mode monitored by a photodetector that encodes the arrival
time $t_{i}$ of each photon in its output record $N\left(t\right)$.
(b) The same setup depicted using cQED components. The optical cavity
with asymmetric mirrors is replaced by
a microwave transmission line resonator with asymmetric coupling
capacitors. The free-space links are replaced by transmission lines.
The circulator ensures that the photodetector
causes no direct back-action on the system,
even if the detector is not ideal.
\label{fig:1q-setup} }
\end{figure}

Measurement-induced dephasing and the possibility of reversing it
has been extensively discussed in the context of dispersive qubit
measurement using quadrature detectors \cite{Schuster2005Ac,Gambetta2006Qubitphoton,Korotkov2006Undoing,Gambetta2008Quantum,Lalumiere2010Tunable,Tornberg2010Highfidelity,FriskKockum2012Undoing,Riste2013Deterministic,deLange2014Reversing}.
The principle of reversing the dephasing using a photodetector has
also been described by Frisk Kockum et al. in Ref.~\cite{FriskKockum2012Undoing}.
Nevertheless, we begin by reviewing these concepts in a single-qubit
scenario because it maps one-to-one to the even-parity subspace of
the two-qubit scenario discussed in Sec.~\ref{sec:parity}.

We consider the qubit--resonator system illustrated in Fig.~\ref{fig:1q-setup}.
We assume that the coupling is dispersive so that the closed-system
Hamiltonian in the Schrödinger picture is well approximated by
\begin{align*}
\hat{H}_{\textrm{q-r}} & =\hbar\left(\omega_{\textrm{r}}+\chi\hat{\sigma}_{z}\right)\hat{a}^{\dagger}\hat{a}+\frac{\hbar\tilde{\omega}_{\textrm{a}}}{2}\hat{\sigma}_{z}\\
 & \quad+\hbar\left[\varepsilon\left(t\right)e^{-i\omega_{\textrm{r}}t}\hat{a}^{\dagger}+\varepsilon^{*}\left(t\right)e^{i\omega_{\textrm{r}}t}\hat{a}\right],
\end{align*}
as given in Ref.~\cite{Gambetta2006Qubitphoton}. Here, $\hat{\sigma}_{z}=\ket e\bra e-\ket g\bra g$,
$\hat{a}$ is the annihilation operator for photons in the resonator,
$\omega_{\textrm{r}}/2\pi$ is the resonator frequency, $\tilde{\omega}_{\textrm{a}}/2\pi$
is the Lamb-shifted qubit frequency, $\chi/\pi$ is the dispersive
frequency shift of the qubit per photon, and $\varepsilon\left(t\right)$
describes the amplitude of a classical drive of the resonator through
a weakly-coupled port. However, we work in an interaction picture
where we transform the basis states by $\mbox{exp}\left[i\left(\omega_{\textrm{r}}\hat{a}^{\dagger}\hat{a}+\tilde{\omega}_{\textrm{a}}\hat{\sigma}_{z}/2\right)t\right]$
and correspondingly redefine 
\begin{align}
\hat{H}_{\textrm{q-r}} & =\hbar\chi\hat{\sigma}_{z}\hat{a}^{\dagger}\hat{a}+\hbar\left[\varepsilon\left(t\right)\hat{a}^{\dagger}+\varepsilon^{*}\left(t\right)\hat{a}\right].\label{eq:ham-1q}
\end{align}
 We consider a drive that displaces the resonator from vacuum $\ket 0$
into a coherent state $\ket{\alpha}=\hat{D}\left(\alpha\right)\ket 0$
at $t=0$ but is otherwise off. This is approximately the case for
a short but strong Gaussian pulse $\varepsilon\left(t\right)=i\alpha e^{-t^{2}/2T_{\textrm{d}}^{2}}/\sqrt{2\pi T_{\textrm{d}}^{2}}$,
where $\kappa,\chi\ll T_{\textrm{d}}^{-1}\ll\omega_{\textrm{r}}$.
We note that the average photon number $\left|\alpha\right|^{2}$
in the initial state is typically limited by assumptions underlying
the dispersive coupling approximation \cite{Gambetta2006Qubitphoton}.

We assume that the dominant interaction between the qubit--resonator
system and its environment is a weak linear coupling between the resonator
and a transmission line with a photodetector at
the other end.
We assume that the photodetector emits negligible noise power to its input
and that it is impedance matched, i.e.,
the detector is operated in the scattering mode \cite{Clerk2010Introduction}.
Given
the rotating-wave, secular, and Born--Markov approximations, the corresponding
stochastic master equation for the reduced density operator $\hat{\rho}_{I}\left(t\right)$
of the qubit--resonator system conditioned on the photodetector output
is 
\begin{align}
d\hat{\rho}_{I} & =dN\left(t\right)\mathcal{G}\left[\sqrt{\eta\kappa}\hat{a}\right]\hat{\rho}_{I}\nonumber \\
 & \quad-dt\mathcal{H}\left[\frac{i}{\hbar}\hat{H}_{\textrm{q-r}}+\frac{\eta\kappa}{2}\hat{a}^{\dagger}\hat{a}\right]\hat{\rho}_{I}\nonumber \\
 & \quad+dt\mathcal{D}\left[\sqrt{\left(1-\eta\right)\kappa}\hat{a}\right]\hat{\rho}_{I},\label{eq:1q-sme-general}
\end{align}
as given in Ref.~\cite{Wiseman2009Quantum}. Here, $\eta$ is the
efficiency of the photodetector, $dN\left(t\right)\in\left\{ 0,1\right\} $
encodes the photon arrival times $T=\left\{ t_{i}|dN\left(t_{i}\right)=1\right\} $,
and $\left\langle dN\left(t\right)\right\rangle =dt\,\mbox{tr}\left[\eta\kappa\hat{a}^{\dagger}\hat{a}\hat{\rho}_{I}\right]$
gives the detection probability. The superoperator $\mathcal{H}\left[\hat{c}\right]\hat{\rho}=\hat{c}\hat{\rho}+\hat{\rho}\hat{c}^{\dagger}-\mbox{tr}\left[\hat{c}\hat{\rho}+\hat{\rho}\hat{c}^{\dagger}\right]\hat{\rho}$
describes the continuous evolution between detection events, $\mathcal{G}\left[\hat{c}\right]\hat{\rho}=\hat{c}\hat{\rho}\hat{c}^{\dagger}/\mbox{tr}\left[\hat{c}\hat{\rho}\hat{c}^{\dagger}\right]-\hat{\rho}$
accounts for the discrete jumps at $T$, and $\mathcal{D}\left[\hat{c}\right]\hat{\rho}=\hat{c}\hat{\rho}\hat{c}^{\dagger}-\frac{1}{2}\left(\hat{c}^{\dagger}\hat{c}\hat{\rho}+\hat{\rho}\hat{c}^{\dagger}\hat{c}\right)$
includes the effects of unmonitored decay. The subscript $I$ emphasizes
that the solutions, called trajectories, depend on the stochastic
photodetector output. In general, Eq.~(\ref{eq:1q-sme-general})
should include terms corresponding to other imperfectly monitored
decay channels, such as $\mathcal{D}\left[\sqrt{\mbox{\ensuremath{\gamma}}}\hat{\sigma}_{-}\right]\hat{\rho}_{I}$
describing spontaneous relaxation of the qubit, but we assume that
$\kappa$ is sufficiently large to neglect them.

\subsection{Unmonitored system: measurement-induced dephasing\label{sub:1q-unmonitored}}

\begin{figure}
\includegraphics{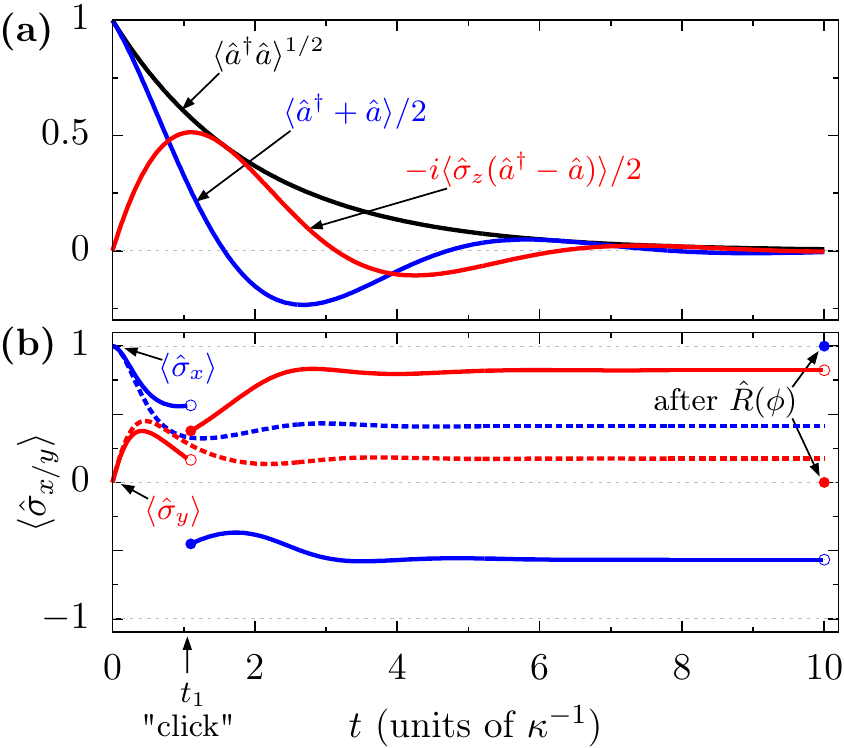}
\caption{(Color online) Example trajectory (solid lines) and average evolution (dashed lines)
of the qubit--resonator system (see Fig.~\ref{fig:1q-setup}) for
$\chi=\kappa$, $\alpha=1$, and the qubit initially in $\left(\ket g+\ket e\right)/\sqrt{2}$.
(a) Expectation values that are independent of the photon arrival
times $T=\left\{ t_{i}\right\} $: square root of the photon number
$\hat{a}^{\dagger}\hat{a}$, the initially excited quadrature $\left(\hat{a}^{\dagger}+\hat{a}\right)/2$,
and product of the conjugate quadrature $i\left(\hat{a}^{\dagger}-\hat{a}\right)/2$
and the qubit operator $-\hat{\sigma}_{z}$. Individually, $\left\langle \hat{a}^{\dagger}-\hat{a}\right\rangle =\left\langle \hat{\sigma}_{z}\right\rangle =0$
for all $t$ (not shown). (b) Expectation values of $\hat{\sigma}_{x}$
and $\hat{\sigma}_{y}$ which show decoherence of the qubit in the
computational basis. The solid points at $t=10\kappa^{-1}$ show the
state after applying a phase gate $\hat{R}\left(\phi\right)$ that
reverses the measurement-induced dephasing $\phi\left(T\right)$ {[}see
Eq.~(\ref{eq:1q-phi}){]}. The curves describe numerical solutions to
Eqs.~(\ref{eq:1q-me}) and~(\ref{eq:1q-sme}).\label{fig:1q-trajectories} }
\end{figure}

In the limit $\eta\rightarrow0$, Eq.~(\ref{eq:1q-sme-general})
reduces to the deterministic master equation
\begin{equation}
\partial_{t}\hat{\rho}=-\frac{i}{\hbar}\left[\hat{H}_{\textrm{q-r}},\hat{\rho}\right]+\mathcal{D}\left[\sqrt{\kappa}\hat{a}\right]\hat{\rho},\label{eq:1q-me}
\end{equation}
where we have dropped the subscript $I$ since the evolution is independent
of the photodetector output. Furthermore, in the scattering mode the
detector type is irrelevant for determining $\hat{\rho}$, and hence
Eq.~(\ref{eq:1q-me}) is the same as for quadrature detection \cite{Gambetta2006Qubitphoton,Wiseman2009Quantum}.

Gambetta et al. solved Eq.~(\ref{eq:1q-me}) for a general drive
$\varepsilon\left(t\right)$ \cite{Gambetta2006Qubitphoton}. For
the specific parameters in our work,
the solution can be expressed in closed form.
For $t>0$, it is a
superposition of two coherent resonator states entangled with the
qubit state:
\begin{equation}
\hat{\rho}\left(t\right)=\sum_{i,j\in\left\{ e,g\right\} }c_{ij}\left(t\right)\ket i\bra j\otimes\ket{\alpha_{i}\left(t\right)}\bra{\alpha_{j}\left(t\right)},\label{eq:1q-rho-form}
\end{equation}
where $\alpha_{g}=\alpha e^{\left(i\chi-\kappa/2\right)t}$ ($\alpha_{e}=\alpha e^{\left(-i\chi-\kappa/2\right)t}$)
describes the exponentially decaying coherent resonator state given
that the qubit is in $\ket g$ ($\ket e$) {[}see Fig.~\ref{fig:1q-trajectories}(a){]}.
The diagonal elements $c_{gg}$ and $c_{ee}$ remain at their initial
values since there are no terms in Eq.~(\ref{eq:1q-me}) that flip
$\ket g$ and $\ket e$. The off-diagonal elements decay according
to $c_{eg}\left(t\right)=a_{eg}\left(t\right)/\braket{\alpha_{g}\left(t\right)}{\alpha_{e}\left(t\right)}$
and $c_{eg}\left(t\right)=c_{ge}^{*}\left(t\right)$, where $\braket{\alpha_{g}\left(t\right)}{\alpha_{e}\left(t\right)}=\mbox{exp}\left[-\left|\alpha\right|^{2}e^{-\kappa t}\left(1-e^{2i\chi t}\right)\right]$,
\begin{align}
a_{eg}\left(t\right) & =c_{eg}\left(0\right)\mbox{exp}\left[-2i\chi\int_{0}^{t}\alpha_{e}\left(t^{\prime}\right)\alpha_{g}^{*}\left(t^{\prime}\right)dt^{\prime}\right]\nonumber \\
 & =c_{eg}\left(0\right)\mbox{exp}\left[-\left|\alpha\right|^{2}\frac{1-e^{\left(2i\chi-\kappa\right)t}}{1-i\kappa/2\chi}\right],\label{eq:a_eg}
\end{align}
and $c_{eg}\left(0\right)$ is the initial value of the off-diagonal
element.

Here $\left|c_{eg}\left(t\right)\right|$ quantifies the phase coherence
remaining in the qubit--resonator system as a whole, while $2a_{eg}\left(t\right)=\left\langle \hat{\sigma}_{x}\left(t\right)\right\rangle -i\left\langle \hat{\sigma}_{y}\left(t\right)\right\rangle $
describes the lateral components of the qubit Bloch vector after tracing
out the resonator {[}see Fig.~\ref{fig:1q-trajectories}(b){]}. For
$t\gg\kappa^{-1}$, the two are equivalent because the resonator state
approaches vacuum exponentially irrespective of the qubit state. Furthermore,
in the long time limit $c_{eg}\left(t\right)/c_{eg}\left(0\right)\approx\mbox{exp}\left[-\left|\alpha\right|^{2}/\left(1-i\kappa/2\chi\right)\right]$
so the net effect of the process on the qubit consists of a coherent
rotation by an angle 
\begin{equation}
\phi_{\textnormal{uncond}}=-\left|\alpha\right|^{2}/\left(2\chi/\kappa+\kappa/2\chi\right)\label{eq:phi-uncond}
\end{equation}
  together with a reduction of coherence by a factor of $\mbox{exp}\left[-\left|\alpha\right|^{2}/\left(1+\kappa^{2}/4\chi^{2}\right)\right]$.
The latter is called measurement-induced dephasing because it is directly
related to the ability to determine the qubit state by monitoring
the radiation leaking out of the system \cite{Gambetta2006Qubitphoton,Gambetta2008Quantum}.

\subsection{Reversing dephasing using a photodetector\label{sub:1q-monitored}}

In the opposite limit of perfect photodetection ($\eta\rightarrow1$),
Eq.~(\ref{eq:1q-sme-general}) reduces to
\begin{align}
d\hat{\rho}_{I} & =dN\left(t\right)\mathcal{G}\left[\sqrt{\kappa}\hat{a}\right]\hat{\rho}_{I}\nonumber \\
 & \quad-dt\mathcal{H}\left[\frac{i}{\hbar}\hat{H}_{\textrm{q-r}}+\frac{\kappa}{2}\hat{a}^{\dagger}\hat{a}\right]\hat{\rho}_{I},\label{eq:1q-sme}
\end{align}
which does not include any unmonitored decay channels and therefore
does not change the purity of the initial state \cite{Wiseman2009Quantum}.
Since the purity does not change and there are no terms that flip
$\ket g$ and $\ket e$, the initial qubit state must be restored
in the long-time limit by a phase gate 
\[
\hat{R}\left(\phi\right)=e^{-i\phi/2}\ket g\bra g+e^{i\phi/2}\ket e\bra e,
\]
where $\phi$ is determined by solving Eq.~(\ref{eq:1q-sme}) for
a given $dN\left(t\right)$.
Below, we show that this stochastic phase is
independent of the initial qubit state and can be written in closed
form as
\begin{equation}
\phi\left(T\right)=2\chi\sum_{t_{i}\in T}t_{i}.\label{eq:1q-phi}
\end{equation}

Equation~(\ref{eq:1q-phi}) can be intuitively understood by noting
that the photons are all injected into the resonator at $t=0$, and
hence the time each detected photon interacts with the qubit is equal
to $t_{i}$. The contributions add linearly so the total accumulated
dispersive phase shift between $\ket e$ and $\ket g$ is $\sum_{i}2\chi t_{i}$,
as shown below for a pure initial state. Frisk Kockum et al. used
an alternative method of solving the problem by applying a polaron
transformation that allows writing a stochastic master equation for
a two-level system only~\cite{FriskKockum2012Undoing}. Specifically,
choosing $\eta=1$, $\epsilon_{m}t_{\textnormal{meas}}=i\alpha$,
and $t_{\textnormal{meas}}\ll\chi^{-1},\kappa^{-1}$ for the parameters
defined in Ref.~\cite{FriskKockum2012Undoing} corresponds to the
scenario we consider. Note, however, that the analysis in Ref.~\cite{FriskKockum2012Undoing}
excludes the coherent evolution due to the ac Stark shift $2\chi\textnormal{Re}\left(\alpha_{g}\alpha_{e}^{\star}\right)$.
Compared to our results, this leads to an additional phase factor
that approaches $\mbox{exp}\left(-i\phi_{\textnormal{uncond}}\right)$
for $t\gg\kappa^{-1}$ (see Eq.~(26) in Ref.~\cite{FriskKockum2012Undoing}).

For a pure initial state, Eq.~(\ref{eq:1q-sme}) for the density
operator $\hat{\rho}_{I}=\ket{\psi_{I}}\bra{\psi_{I}}$ is equivalent
to a stochastic Schrödinger equation for an unnormalized state $\ket{\psi_{I}^{\prime}\left(t\right)}=\braket{\psi_{I}^{\prime}\left(t\right)}{\psi_{I}^{\prime}\left(t\right)}^{1/2}\ket{\psi_{I}}$
\cite{Wiseman2009Quantum}. For $t>0$, this means solving the Schrödinger
equation for a non-Hermitian Hamiltonian $\hbar\left(\chi\hat{\sigma}_{z}-i\kappa/2\right)\hat{a}^{\dagger}\hat{a}$
between detection events, while a detected photon is taken into account
by applying the jump operator $\hat{a}$ and renormalizing the resulting
state. A photon is detected whenever $\braket{\psi_{I}^{\prime}\left(t\right)}{\psi_{I}^{\prime}\left(t\right)}$
reaches a random number $r_{i}$ drawn uniformly and independently
from $\left[0,1\right]$ for each detection event $i$.

For a normalized initial state $\left(q_{g}\ket g+q_{e}\ket e\right)\ket{\alpha}$,
the unnormalized state before the first detection event is
\begin{align*}
\ket{\psi_{I}^{\prime}\left(t\right)} & =e^{-\left|\alpha\right|^{2}\left(1-e^{-\kappa t}\right)/2}\left(q_{g}\bket g\bket{\alpha e^{\left(i\chi-\kappa/2\right)t}}\right.\\
 & \quad\quad\quad\quad\quad\quad\quad\quad\,\,\,+\left.q_{e}\bket e\bket{\alpha e^{\left(-i\chi-\kappa/2\right)t}}\right).
\end{align*}
If $r_{1}>e^{-\left|\alpha\right|^{2}}$, a photon is detected at
$t_{1}=-\kappa^{-1}\ln\left[1+\left|\alpha\right|^{-2}\ln\left(r_{1}\right)\right]$
and the normalized state after applying $\hat{a}$ becomes
\[
q_{g}e^{i\chi t_{1}}\bket g\bket{\alpha e^{\left(i\chi-\kappa/2\right)t}}+q_{e}e^{-i\chi t_{1}}\bket e\bket{\alpha e^{\left(-i\chi-\kappa/2\right)t}}.
\]
Similarly, each subsequent detection event results in factors of $e^{\pm i\chi t_{i}}$,
leading to the normalized state 
\begin{align*}
\ket{\psi_{I}\left(t\right)} & =q_{g}\exp\left[i\chi\sum_{t_{i}<t}t_{i}\right]\ket g\ket{\alpha e^{(i\chi-\kappa/2)t}}\\
 & \quad+q_{e}\exp\left[-i\chi\sum_{t_{i}<t}t_{i}\right]\ket e\ket{\alpha e^{(-i\chi-\kappa/2)t}},
\end{align*}
where $t_{i}\in T$ and no more jumps occur when $r_{i}<\exp\left(-\left|\alpha\right|^{2}e^{-\kappa t_{i-1}}\right)$,
with $t_{0}=0$. See Fig.~\ref{fig:1q-trajectories}(b) for an example
with one detection event.

Evidently applying $\hat{R}\left[\phi\left(T|t_{i}<t\right)\right]$
to $\ket{\psi_{I}\left(t\right)}$ undoes the relative phase between
the $\ket g$ and $\ket e$ terms above. In the long-time limit $t\gg\kappa^{-1}\log\left|\alpha\right|$,
the overlap $\left|\matrixel{\psi_{I}\left(0\right)}{\hat{R}\left(\phi\right)}{\psi_{I}\left(t\right)}\right|$
furthermore approaches unity and additional detection events become
exponentially unlikely as the resonator returns to $\ket 0$. Therefore
we conclude that the combination of photodetection and $\hat{R}\left(\phi\right)$
effectively erases all the information that leaked out of the qubit
during the process. In the parity measurement protocol, the same principle
is used to erase only the part of the information that would allow
distinguishing between $\ket{gg}$ and $\ket{ee}$ if a different
scattering-mode detection scheme were used.

Note that the resonator and the qubit periodically return to a product
state, i.e., $\alpha_{g}\left(\tilde{t}_{k}\right)=\alpha_{e}\left(\tilde{t}_{k}\right)=\left(-1\right)^{k}e^{-\kappa\tilde{t}_{k}/2}$
for all $\tilde{t}_{k}\in\left\{ k\pi\chi^{-1}|k\in\mathbb{Z}^{+}\right\} $
{[}see Fig.~\ref{fig:1q-trajectories}(a){]}. At these times the
dynamics can be stopped by a second displacement $\hat{D}\left[-\alpha_{g}\left(\tilde{t}_{k}\right)\right]$
that brings the resonator to $\ket 0$ deterministically, thereby
decoupling the duration of the measurement from $\kappa^{-1}$ in
the $\chi\apprge\kappa$ regime. We take advantage of this possibility
in the parity measurement protocol proposed in the next section.

\section{Remote Parity Measurement\label{sec:parity}}

\begin{figure}
\includegraphics{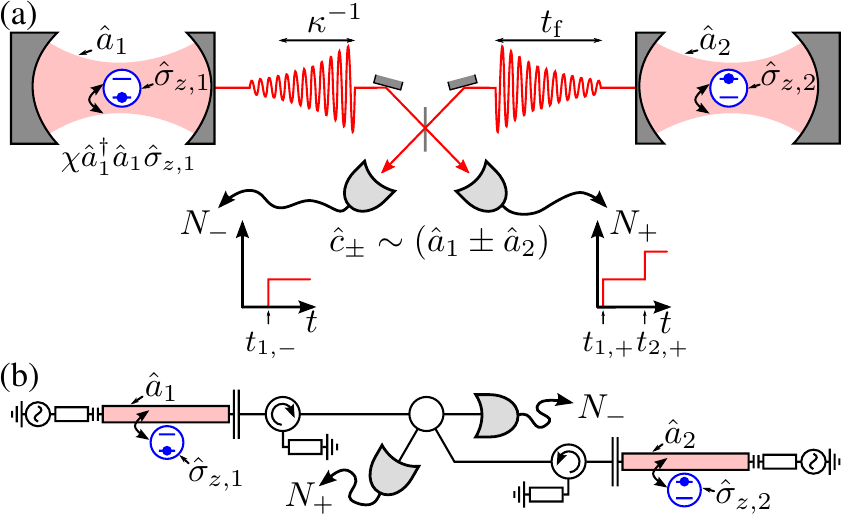}
\caption{(Color online) (a) Schematic setup for measuring the parity $\hat{\sigma}_{z,1}\hat{\sigma}_{z,2}$
of two qubits located in identical resonators (see also Fig.~\ref{fig:1q-setup}).
The resonators are driven into a coherent state $\ket{\alpha}$ at
$t=0$ after which their phase becomes entangled with the local qubit
at a rate $\chi$. At a rate $\kappa$, the resonators radiate into
a 50:50 beam splitter with photodetectors monitoring the output modes.
The relative phases are chosen such that the detectors monitor $\hat{c}_{\pm}=\left(\hat{a}_{1}\pm\hat{a}_{2}\right)/\sqrt{2}$.
At a final time $t_{\textrm{f}}$ (integer multiple of $\pi\chi^{-1}$)
the resonators are displaced back to vacuum. Recording more than zero
photons for $\hat{c}_{-}$ {[}$N_{-}\left(t_{\textrm{f}}\right)>0${]}
indicates odd parity with certainty while $N_{-}\left(t_{\textrm{f}}\right)=0$
partially projects the qubits to the even-parity subspace.
The final
$N_{-}\left(t_{\textrm{f}}\right)$
is sufficient for reversing measurement-induced dephasing within the
odd-parity subspace but the full time-resolved $N_{+}\left(t\right)$
is needed in the even-parity case.
(b) The same setup depicted using cQED components.
The beam splitter is replaced by a $180^{\circ}$ hybrid coupler.
\label{fig:2q-setup}}
\end{figure}

Figure~\ref{fig:2q-setup} shows the setup we propose for measuring
the parity $\hat{\sigma}_{z,1}\hat{\sigma}_{z,2}$ of two remote qubits.
It consists of two identical instances of the dispersively coupled
qubit--resonator system described in the previous section. The two
resonator modes $\hat{a}_{1}$ and $\hat{a}_{2}$ are driven to an
identical coherent state $\ket{\alpha}$ at $t=0$ but the radiation
leaking out of them is not measured individually. Rather, a 50:50
beam splitter is arranged such that two identical photodetectors monitor
the sum and difference modes with lowering operators $\hat{c}_{\pm}=\left(\hat{a}_{1}\pm\hat{a}_{2}\right)/\sqrt{2}$.
The stochastic master equation for this system is
\begin{align}
d\hat{\rho}_{I} & =dN_{+}\left(t\right)\mathcal{G}\left[\sqrt{\eta\kappa}\hat{c}_{+}\right]\hat{\rho}_{I}\nonumber \\
 & \quad+dN_{-}\left(t\right)\mathcal{G}\left[\sqrt{\eta\kappa}\hat{c}_{-}\right]\hat{\rho}_{I}\nonumber \\
 & \quad-dt\mathcal{H}\left[\frac{i}{\hbar}\hat{H}_{\textrm{q-r}}^{(2)}+\frac{\eta\kappa}{2}\hat{c}_{+}^{\dagger}\hat{c}_{+}+\frac{\eta\kappa}{2}\hat{c}_{-}^{\dagger}\hat{c}_{-}\right]\hat{\rho}_{I}\nonumber \\
 & \quad+dt\mathcal{D}\left[\sqrt{\left(1-\eta\right)\kappa}\hat{c}_{+}\right]\hat{\rho}_{I}\nonumber \\
 & \quad+dt\mathcal{D}\left[\sqrt{\left(1-\eta\right)\kappa}\hat{c}_{-}\right]\hat{\rho}_{I},\label{eq:2q-sme}
\end{align}
where $\eta$ is the photodetection efficiency, $N_{\pm}\left(t\right)$
is the number of photons registered by the detector monitoring $\hat{c}_{\pm}$,
and
\begin{equation}
\hat{H}_{\textrm{q-r}}^{(2)}=\hbar\chi\left(\hat{\sigma}_{z,1}\hat{a}_{1}^{\dagger}\hat{a}_{1}+\hat{\sigma}_{z,2}\hat{a}_{2}^{\dagger}\hat{a}_{2}\right)\label{eq:ham_2q}
\end{equation}
for $t>0$. We denote coherent states of the sum and difference modes
by $\ket{\beta}_{\pm}$.

The parity measurement and the measurement-induced dephasing in this
scenario are closely connected to the case of multiple qubits coupled
to the same resonator mode \cite{Hutchison2009Quantum,Bishop2009Proposal,Helmer2009Measurementbased,Lalumiere2010Tunable,Tornberg2010Highfidelity,Riste2013Deterministic,FriskKockum2012Undoing,Nigg2013Stabilizer,Govia2015Scalable}.
Within the even-parity subspace, the difference mode in fact remains
in vacuum so, mathematically, the description of the system maps one-to-one
to the single resonator case. However, physically the situation is
distinct because the populated resonator mode is the non-local sum
mode. In the odd-parity subspace neither mode remains in vacuum.

\subsection{Protocol}

In this section,
we describe the proposed RNPM protocol and its effect on the
qubits. Section~\ref{sub:2q-sse} proves the validity of these claims
analytically for $\eta\rightarrow1$. Additionally, Appendix~\ref{app:photon-loss}
presents some numerical trajectories in non-ideal cases.

The steps of the protocol are the following:
\begin{enumerate}
\item Start with the resonators in $\ket 0$ and the qubits in an arbitrary
pure state
\begin{equation}
\ket{\psi_{\textrm{q}}\left(0\right)}=\sum_{i,j\in\left\{ e,g\right\} }q_{ij}\ket{ij}.\label{eq:2q-initial}
\end{equation}
\item Apply $\hat{D}\left(\alpha\right)$ to both resonators at $t=0$.\label{enu:disp-start}
\item Wait until $t_{\textrm{f}}=k\pi\chi^{-1}$, where $k\in\mathbb{\mathbb{Z}}^{+}$
is arbitrary.
\item Apply $\hat{D}\left[-\left(-1\right)^{k}\alpha e^{-\kappa t_{\textrm{f}}/2}\right]$
to both resonators.\label{enu:disp-end}
\item Compute the measurement induced phases
\begin{align}
\phi_{+} & =2\chi\sum_{t_{i}\in T_{+}}t_{i}\label{eq:2q:phiplus}\\
\textnormal{and }\phi_{-} & =\pi N_{-}\left(t_{\textrm{f}}\right),\label{eq:2q:phiminus}
\end{align}
 and apply the local qubit feedback operations 
\begin{align}
\hat{F} & =\hat{R}\left(\frac{\phi_{+}}{2}+\frac{\phi_{-}}{2}\right)\otimes\hat{R}\left(\frac{\phi_{+}}{2}-\frac{\phi_{-}}{2}\right).\label{eq:2q:F}
\end{align}
\end{enumerate}
Assuming $\eta\rightarrow1$, the state of the qubits after these
steps is 
\begin{equation}
\ket{\psi_{\textrm{q}}\left(t_{\textrm{f}}\right)}=\left[\sqrt{P_{-1}}\hat{\Pi}_{-1}+\sqrt{P_{1}}\hat{\Pi}_{1}\right]\ket{\psi_{\textrm{q}}\left(0\right)},\label{eq:postmeas-qubits}
\end{equation}
where $P_{1}=1-P_{-1}$. The outcome of the parity measurement is
indicated by $P_{-1}=\expect{\psi_{\textrm{q}}\left(t_{\textrm{f}}\right)}{\hat{\Pi}_{-1}}$
and is determined by $N_{-}\left(t_{\textrm{f}}\right)$ and $T_{+}=\left\{ t_{i}|dN_{+}\left(t_{i}\right)=1\right\} $
according to 
\[
P_{-1}=\begin{cases}
1\quad\mbox{if }N_{-}\left(t_{\textrm{f}}\right)>0\\
\left(\left|q_{ge}\right|^{2}+\left|q_{eg}\right|^{2}\right)\left({\displaystyle \prod_{t_{i}\in T_{+}}\cos\chi t_{i}}\right)^{2} & \mbox{otherwise.}
\end{cases}
\]

Evidently a single $\hat{c}_{-}$ detection event leads to complete
parity projection to the odd-parity subspace ($\left\langle \hat{\sigma}_{z,1}\hat{\sigma}_{z,1}\right\rangle =-1$),
while $\hat{c}_{+}$ detections in general lead to an exponential
suppression of the odd-parity components, as shown in Fig.~\ref{fig:trajectories}.
Therefore the outcome of the protocol is not strictly speaking binary,
but rather continuous in the interval $P_{-1}\in[0,1]$. However,
in the limit of many photodetection events, the protocol is well approximated
by a projective parity measurement with $P_{-1}\in\left\{ 0,1\right\} $,
up to exponentially small corrections given by the exact result. The
expected number of detected photons is $\left|\alpha\right|^{2}\left(1-e^{-\kappa t_{\textrm{f}}}\right)$
and therefore tunable. Furthermore, if $P_{-1}$ is not sufficiently
close to zero or one at $t_{\textrm{f}}$, steps \ref{enu:disp-start}
through \ref{enu:disp-end} of the protocol can be repeated arbitrarily
many times at the expense of increased operation time. In that case,
a single feedback operation $\hat{F}$ applied after the last repetition
should account for all detected photons. The displacements in steps
\ref{enu:disp-end} and \ref{enu:disp-start} of adjacent repetitions
can be combined into a single operation as well.

Finally, we claim that the outcome probabilities are distributed according
to the parity of the initial state. More specifically, the ensemble
average of $P_{-1}$ is $\textnormal{E}\left[P_{-1}\right]=\expect{\psi_{\textrm{q}}\left(0\right)}{\hat{\Pi}_{-1}}$.
This implies that the protocol is indeed a parity measurement, rather
than some other operation that leads to a well-defined parity.

\subsection{Full temporal evolution\label{sub:2q-sse}}

\begin{figure}
\includegraphics{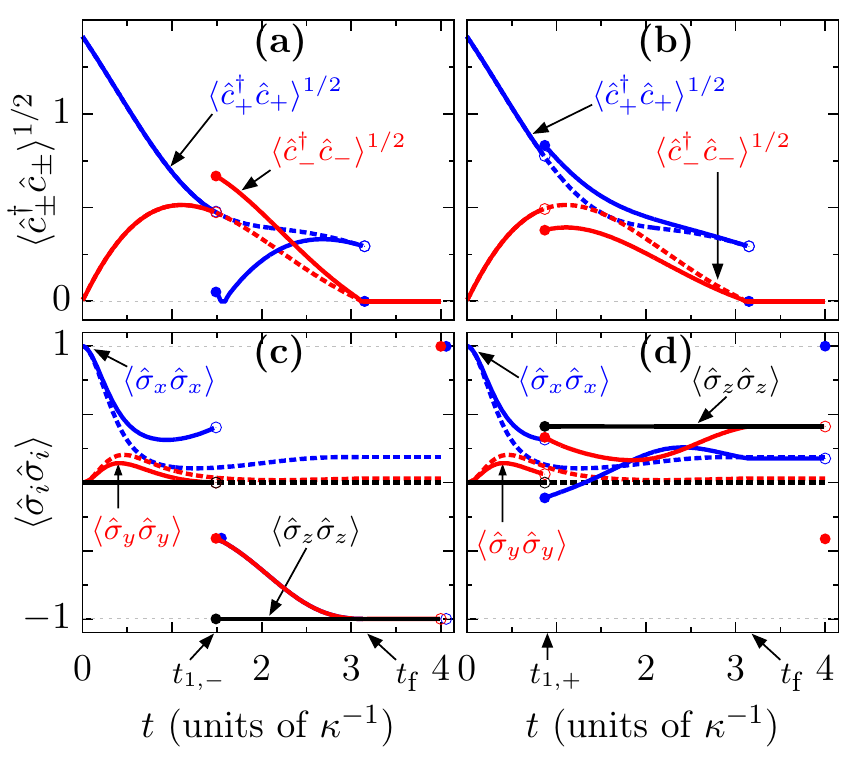}
\caption{(Color online) Example trajectories (solid lines) and average evolution
(dashed lines) for the two-qubit setup shown in Fig.~\ref{fig:2q-setup}
with $\chi=\kappa$, $\alpha=1$, $t_{\textrm{f}}=\pi\chi^{-1}$,
and the qubits initially in $\left(\ket g+\ket e\right)^{\otimes2}/2$.
The left column (a,c) shows a trajectory that leads to projection
of the qubits into the odd-parity subspace, while the right column
(b,d) shows partial projection into the even-parity subspace. (a,b)
The expected photodetection rates in units of $\kappa$ for the sum
and difference modes ($\hat{c}_{\pm}$), with $t_{1,\pm}$ indicating
the first and only photodetection event in the selected trajectories.
Up to $t_{\textrm{f}}$, the local field quadratures $\left\langle \hat{a}_{j}^{\dagger}+\hat{a}_{j}\right\rangle /2$
and $-i\hat{\sigma}_{z,j}\left\langle \hat{a}_{j}^{\dagger}-\hat{a}_{j}\right\rangle /2$
are identical to those shown in Fig.~\ref{fig:1q-trajectories}(a),
regardless of detection events. (c,d) Expectation values of the Bell
state stabilizers $\hat{\sigma}_{i,1}\hat{\sigma}_{i,2}$, where $i\in\left\{ x,y,z\right\} $.
A projective parity measurement of the specified initial state leads
to $\left\langle \hat{\sigma}_{z,1}\hat{\sigma}_{z,2}\right\rangle =\pm1$
and conserves $\left\langle \hat{\sigma}_{x,1}\hat{\sigma}_{x,2}\right\rangle =1$.
The solid points at $t=4\kappa^{-1}$ show the state after applying
a phase gate $\hat{F}$ {[}see Eq.~(\ref{eq:2q:F}){]} that reverses
the measurement-induced dephasing, i.e., restores $\left\langle \hat{\sigma}_{x,1}\hat{\sigma}_{x,2}\right\rangle =1$.
The curves describe numerical solutions to Eqs.~(\ref{eq:2q-sme}) and~(\ref{eq:2q-me})
with $\eta=1$.\label{fig:trajectories} Overlapping points have been
offset horizontally for clarity.}
\end{figure}

Let us first give a qualitative explanation for Eqs.~(\ref{eq:2q:phiplus})--(\ref{eq:postmeas-qubits})
by rewriting the Hamiltonian in Eq.~(\ref{eq:ham_2q}) in the basis
of the monitored operators as
\begin{align*}
\hat{H}_{\textrm{q-r}}^{(2)} & =\frac{\hbar\chi}{2}\left[(\hat{\sigma}_{z,1}+\hat{\sigma}_{z,2})(\hat{c}_{+}^{\dagger}\hat{c}_{+}+\hat{c}_{-}^{\dagger}\hat{c}_{-})\right.\\
 & \qquad\quad\left.+(\hat{\sigma}_{z,1}-\hat{\sigma}_{z,2})(\hat{c}_{+}^{\dagger}\hat{c}_{-}+\hat{c}_{+}\hat{c}_{-}^{\dagger})\right].
\end{align*}
Note that after the initial displacements the sum and difference modes
start in $\ket{\sqrt{2}\alpha}_{+}$ and $\ket 0_{-}$, respectively.
Furthermore, there are no terms in Eq.~(\ref{eq:2q-sme}) that flip
the qubits in the computational basis so parity is conserved if it
is initially well defined. For an even-parity initial state, $\hat{\sigma}_{z,1}-\hat{\sigma}_{z,2}$
yields zero so there are no terms in Eq.~(\ref{eq:2q-sme}) that
excite the difference mode out of $\ket 0_{-}$ and we can therefore
trace it out without loss of information. The remaining terms are
identical to the single qubit case {[}see Eq.~(\ref{eq:1q-sme}){]}
with the mapping $\ket{gg}\rightarrow\ket g$, $\ket{ee}\rightarrow\ket e$,
and $\hat{c}_{+}\rightarrow\hat{a}$. This explains why Eq.~(\ref{eq:2q:phiplus})
matches Eq.~(\ref{eq:1q-phi}). On the other hand for an odd-parity
initial state, $\hat{\sigma}_{z,1}+\hat{\sigma}_{z,2}$ yields zero
so the dispersive $\hat{c}_{\pm}^{\dagger}\hat{c}_{\pm}$ terms in
$\hat{H}_{\textrm{q-r}}^{(2)}$ vanish.
Instead, photons
are exchanged between the two bosonic modes by the $\hat{c}_{\pm}^{\dagger}\hat{c}_{\mp}$
terms with a phase flip between $\ket{ge}$ and $\ket{eg}$ associated
with each exchange. Since the only other event that changes the photon
number in $\hat{c}_{-}$ is a detection event $dN_{-}=1$, the parity
of $N_{-}\left(t_{\textrm{f}}\right)$ and the parity of the number
of phase flips must match given that $\hat{c}_{-}$ starts and ends
in vacuum and $\eta=1$. This explains Eq.~(\ref{eq:2q:phiminus}).

Let us formally prove the validity of Eqs.~(\ref{eq:2q:phiplus})--(\ref{eq:postmeas-qubits})
and the associated probabilities for a pure initial state by explicitly
solving the stochastic Schrödinger equation corresponding to Eq.~(\ref{eq:2q-sme})
in the $\eta\rightarrow1$ limit. Analogously to the single-qubit
case, the jump operators are $\hat{c}_{\pm}$ and the non-Hermitian
Hamiltonian that determines the evolution of the unnormalized state
$\ket{\psi_{I}^{\prime}\left(t\right)}$ is 
\begin{align*}
H^{\prime} & =\hat{H}_{\textrm{q-r}}^{(2)}-i\hbar\frac{\kappa}{2}\left(\hat{c}_{+}^{\dagger}\hat{c}_{+}+\hat{c}_{-}^{\dagger}\hat{c}_{-}\right)\\
 & =\sum_{j=1}^{2}\hbar\left(\chi\hat{\sigma}_{z,j}-i\hbar\frac{\kappa}{2}\right)\hat{a}_{j}^{\dagger}\hat{a}_{j}
\end{align*}
for $t>0$. A photon is detected whenever $\braket{\psi_{I}^{\prime}\left(t\right)}{\psi_{I}^{\prime}\left(t\right)}$
reaches a random number $r_{i}$ drawn uniformly and independently
from $\left[0,1\right]$ for each event $i$. The detector that clicks
is chosen at random with probabilities weighted by $\bexpect{\psi_{I}^{\prime}\left(t\right)}{\hat{c}_{\pm}^{\dagger}\hat{c}_{\pm}}$.

For the initial state given in Eq.~(\ref{eq:2q-initial}), the unnormalized
state $\ket{\psi_{I}^{\prime}\left(t\right)}$ before the first detection
event is
\begin{align*}
e^{-\left|\alpha\right|^{2}\left(1-e^{-\kappa t}\right)} & \left(q_{gg}\bket{gg}\bket{Ae^{i\chi t}}_{+}\bket 0_{-}\right.\\
 & \,\,+q_{ee}\bket{ee}\bket{Ae^{-i\chi t}}_{+}\bket 0_{-}\\
 & \,\,+q_{ge}\bket{ge}\bket{A\cos\chi t}_{+}\bket{iA\sin\chi t}_{-}\\
 & \,\,+\left.q_{eg}\bket{eg}\bket{A\cos\chi t}_{+}\bket{-iA\sin\chi t}_{-}\right),
\end{align*}
where $A\left(t\right)=\sqrt{2}\alpha e^{-\kappa t/2}$ is an exponentially
decaying amplitude of the remaining radiation. If $r_{1}>\exp\left[-2\left|\alpha\right|^{2}\left(1-e^{-\kappa t_{\textrm{f}}}\right)\right]$,
a photon is detected at $t_{1}=-\kappa^{-1}\ln\left[1+\left|\alpha\right|^{-2}\ln\left(r_{1}\right)/2\right]$
and the state $\ket{\psi_{I}\left(t\right)}$ after the event is proportional
to 
\begin{align*}
 & iq_{ge}\bket{ge}\bket{A\cos\chi t}_{+}\bket{iA\sin\chi t}_{-}\\
+ & \left(-i\right)q_{eg}\bket{eg}\bket{A\cos\chi t}_{+}\bket{-iA\sin\chi t}_{-}
\end{align*}
if $dN_{-}\left(t_{1}\right)=1$, or
\begin{align*}
 & e^{i\chi t_{1}}q_{gg}\bket{gg}\bket{Ae^{i\chi t}}_{+}\bket 0_{-}\\
+ & e^{-i\chi t_{1}}q_{ee}\bket{ee}\bket{Ae^{-i\chi t}}_{+}\bket 0_{-}\\
+ & \cos\left(\chi t_{1}\right)q_{ge}\bket{ge}\bket{A\cos\chi t}_{+}\bket{iA\sin\chi t}_{-}\\
+ & \cos\left(\chi t_{1}\right)q_{eg}\bket{eg}\bket{A\cos\chi t}_{+}\bket{-iA\sin\chi t}_{-}
\end{align*}
if $dN_{+}\left(t_{1}\right)=1$. See Fig.~\ref{fig:trajectories}
for examples. The calculation for each subsequent event at $t_{i}$
is identical given the new initial state at $t_{i-1}$. Therefore,
at an arbitrary time $t<t_{\textnormal{f}}$, the state is
\begin{align*}
\ket{\psi_{I}\left(t\right)} & \propto i^{N_{-}\left(t\right)}q_{ge}\bket{ge}\bket{A\cos\chi t}_{+}\bket{iA\sin\chi t}_{-}\\
 & \quad+\left(-i\right)^{N_{-}\left(t\right)}q_{eg}\bket{eg}\bket{A\cos\chi t}_{+}\bket{-iA\sin\chi t}_{-}
\end{align*}
if $N_{-}\left(t\right)>0$, and otherwise
\begin{align*}
\ket{\psi_{I}\left(t\right)} & \propto\exp\left(i\chi\sum_{t_{i}<t}t_{i}\right)q_{gg}\bket{gg}\bket{Ae^{i\chi t}}_{+}\bket 0_{-}\\
 & \quad+\exp\left(-i\chi\sum_{t_{i}<t}t_{i}\right)q_{ee}\bket{ee}\bket{Ae^{-i\chi t}}_{+}\bket 0_{-}\\
 & \quad+\left(\prod_{t_{i}<t}\cos\chi t_{i}\right)\\
 & \quad\quad\times\Bigg(q_{ge}\bket{ge}\bket{A\cos\chi t}_{+}\bket{iA\sin\chi t}_{-}\\
 & \quad\quad\quad\quad+q_{eg}\bket{eg}\bket{A\cos\chi t}_{+}\bket{-iA\sin\chi t}_{-}\Bigg),
\end{align*}
where $t_{i}\in T_{+}$. At $t_{\textrm{f}}=k\pi\chi^{-1}$, the difference
mode returns to vacuum and the sum mode is in $\bket{\left(-1\right)^{k}A\left(t\right)}_{+}$.
The sum mode is then driven back to vacuum by the symmetric displacements
in step \ref{enu:disp-end} of the protocol. Therefore the dynamics
stop and the resonators can be traced out without loss of information.

In the $N_{-}\left(t_{\textrm{f}}\right)>0$ case, the contribution
of $\phi_{-}$ in $\hat{F}$ {[}see Eq.~(\ref{eq:2q:F}){]} undoes
the relative $\left(-1\right)^{N_{-}\left(t_{\textrm{f}}\right)}$
factor between $\ket{ge}$ and $\ket{eg}$, while the contribution
due to $\phi_{+}$ does nothing. Therefore $\ket{\psi_{\textrm{q}}\left(t_{\textrm{f}}\right)}=\hat{F}\left[\braket{0_{+}0_{-}}{\psi_{I}\left(t_{\textrm{f}}\right)}\right]$
indeed matches Eq.~(\ref{eq:postmeas-qubits}) with $P_{-1}=1$.
In the $N_{-}\left(t_{\textrm{f}}\right)=0$ case, the contribution
of $\phi_{+}$ in $\hat{F}$ restores the initial relative phase in
the even-parity subspace so that
\begin{align*}
\ket{\psi_{\textrm{q}}\left(t_{\textrm{f}}\right)} & \propto q_{gg}\bket{gg}+q_{ee}\bket{ee}\\
 & \quad+\left(\prod_{t_{i}<t}\cos\chi t_{i}\right)\left(q_{ge}\bket{ge}+q_{eg}\bket{eg}\right).
\end{align*}
 Note that in the limit $N_{+}\left(t_{\textrm{f}}\right)\gg1$, the
product of the cosines in general vanishes exponentially.

To show that $\textnormal{E}\left[P_{-1}\right]=\bexpect{\psi_{\textrm{q}}\left(0\right)}{\hat{\Pi}_{-1}}$
we use the fact that the expectation value of $\hat{\Pi}_{-1}$ for
the unconditioned system state $\hat{\rho}$ is invariant in time.
The time invariance follows from the fact that the master equation
\begin{align}
\partial_{t}\hat{\rho} & =-\frac{i}{\hbar}\left[\hat{H}_{\textrm{q-r}}^{(2)},\hat{\rho}\right]+\sum_{\pm}\mathcal{D}\left[\sqrt{\kappa}\hat{c}_{\pm}\right]\hat{\rho}\label{eq:2q-me}\\
 & =-\frac{i}{\hbar}\left[\hat{H}_{\textrm{q-r}}^{(2)},\hat{\rho}\right]+\sum_{j=1}^{2}\mathcal{D}\left[\sqrt{\kappa}\hat{a}_{j}\right]\hat{\rho}\nonumber 
\end{align}
does not depend on the type of scattering-mode detection and is solved
by $\hat{\rho}=\hat{\rho}_{1q}\otimes\hat{\rho}_{1q}$, where $\hat{\rho}_{1q}\left(t\right)$
is the solution to Eq.~(\ref{eq:1q-me}) given in Eqs.~(\ref{eq:1q-rho-form})
and (\ref{eq:a_eg}). As noted in Sec.~\ref{sub:1q-unmonitored},
the diagonal elements of $\hat{\rho}_{1q}\left(t\right)$ do not change
in the computational basis. Therefore the diagonal elements of the
two-qubit $\hat{\rho}\left(t\right)$ also remain constant, and hence
$\mbox{tr}\left[\hat{\rho}\left(t\right)\hat{\Pi}_{-1}\right]$ is
time invariant. Applying this time invariance to $\hat{\rho}\left(0\right)=\ket{\psi_{\textrm{q}}\left(0\right)}\bra{\psi_{\textrm{q}}\left(0\right)}$
shows that $\bexpect{\psi_{\textrm{q}}\left(0\right)}{\hat{\Pi}_{-1}}=\mbox{tr}\left[\hat{\rho}\left(0\right)\hat{\Pi}_{-1}\right]$
is equal to $\mbox{tr}\left[\hat{\rho}\left(t_{\textrm{f}}\right)\hat{\Pi}_{-1}\right]=\textnormal{E}\left[P_{-1}\right]$.

\subsection{Tunable coupling\label{sub:dyn-decoup}}

Here we propose a variation of the protocol for a scenario where the
dispersive coupling can be effectively switched on and off in time,
either by changing $\chi$ or by using dynamical decoupling \cite{Viola1999Dynamical}.
If this capability is used to turn off the unitary evolution ($\hat{H}_{\textrm{q-r}}^{(2)}\rightarrow0$)
at $t_{\textrm{off}}=\pi/2\chi$, the results of the previous section
are valid with the replacement of $\chi t$ by $\chi\min\left(t,t_{\textrm{off}}\right)$.
Consequently, a single detection event after $t_{\textrm{off}}$ leads
to complete parity projection because odd (even) qubit parity is associated
with the sum (difference) mode being in vacuum. In this modified protocol
we also require $t_{\textrm{f}}\gg\kappa^{-1}$ and skip step \ref{enu:disp-end}
of the protocol.

This variation of the protocol is particularly beneficial in the strong
dispersive limit \cite{Schuster2007Resolving} $\chi\gg\kappa$ where
most photons leak out of the resonators after $t_{\textrm{off}}$.
In addition to complete parity projection, the variation may be of
practical benefit because $\phi_{+}$ {[}see Eq.~(\ref{eq:2q:phiplus}){]}
becomes independent of the arrival times of the detected photons if
they can be assumed to all arrive after $t_{\textrm{off}}$, i.e.,
if $\chi\gg\left|\mbox{\ensuremath{\alpha}}\right|^{2}\kappa$. In
the extreme limit $\kappa/\chi\rightarrow0$, the initial time interval
up to $t_{\textrm{off}}$ can be approximated as an instantaneous
entanglement operation between each qubit and its resonator.
In this
special case, our protocol after $t_{\textrm{off}}$ coincides with
the protocol proposed by Azuma et al. for the initial state $\theta=\pi$,
where $\theta$ is defined in Ref. \cite{Azuma2012Quantum}.

A detection event at $t>t_{\textrm{off}}$ also projects the resonator
into a known coherent state, so in principle the requirement $t_{\textrm{f}}\gg\kappa^{-1}$
can be relaxed by driving the resonators into vacuum by displacements
conditioned on $N_{\pm}$. Alternatively, $\chi$ can be restored
to its initial value at some time $t^{\prime}$ so that the difference
mode evolves back to vacuum at $t^{\prime}+t_{\textrm{off}}$ regardless
of the qubit state. At $t^{\prime}+t_{\textrm{off}}$ unconditioned
displacements on $\hat{c}_{+}$ can then stop the process analogously
to step~\ref{enu:disp-end} of the original protocol.

\subsection{Practical considerations}

The main experimental hurdle to implementing our protocol in cQED is that
it requires nearly ideal photodetectors. Specifically, in order for
the measurement to be non-destructive the photodetectors need to have
high quantum efficiency, low dark-count rate, photon number resolution
and, for the even-parity outcome, high temporal resolution compared
to $\chi^{-1}$. Here we use number resolution to mean that the detector
must not have a significant dead time after a detection event. By
high quantum efficiency we mean that, in addition to high detector
efficiency, photon losses in other parts of the setup must be negligible.
Failing to satisfy any of these requirements leads to erroneous terms
in Eq.~(\ref{eq:2q:phiplus}) and therefore randomizes the relative
phase within the parity subspaces. Appendix~\ref{app:photon-loss}
shows the effect of imperfect quantum efficiency on example trajectories.

It is possible to relax some of the above requirements: Temporal resolution
is unnecessary in the $\chi\gg\kappa$ case if using the dynamical
decoupling discussed in Sec. \ref{sub:dyn-decoup}. Number resolution
becomes unnecessary if $\left|\alpha\right|^{2}\ll1$ and the protocol
is instead iterated many times. However, the latter increases the
duration of the protocol and will eventually invalidate the assumption
that other relaxation mechanisms are negligible. For reference, $\chi/2\pi$
in cQED is often several megahertz so a single iteration may take
$t_{\textrm{f}}\sim100\mbox{ ns}$. This should be compared to qubit
coherence times that have recently reached roughly $100\,\mbox{\ensuremath{\mu}s}$
\cite{Paik2011Observation,Rigetti2012Superconducting,Barends2013Coherent}.
Appendix~\ref{app:photon-loss} presents some example trajectories
for non-negligible qubit relaxation.

The assumption of identical qubit--resonator systems is another source
of concern for practical implementations. In cQED, $\chi$ is usually
tunable through the qubit frequency but, typically, the resonator
frequency $\omega_{\textrm{r}}$ and the decay rate $\kappa$ are
not tunable. Fortunately, the parameters of typical cQED resonators
are highly reproducible~\cite{Goppl2008Coplanar}. Furthermore, in-situ
tuning of both $\omega_{\textrm{r}}$ and $\kappa$ is possible at
the cost of increased complexity \cite{Osborn2007FrequencyTunable,Pierre2014Storage}.
Finally, choosing an asymmetric drive in step \ref{enu:disp-start}
of the protocol can compensate for different values of $\kappa$.
In general, this only adjusts the average number of emitted photons
per resonator and not the time scale of their emission. However, in
the strong dispersive limit it is possible to choose $t_{\textrm{f}}\ll\kappa^{-1}$
and approximate the photon emission rate as constant.

\section{Conclusion\label{sec:conclusion}}

We proposed a protocol for remote non-destructive parity measurement
of two qubits. The protocol is deterministic in the sense that it
leads to complete parity projection with a probability that approaches
unity in the ideal case. Furthermore, it conserves the relative phase
within the parity subspaces even when the parity projection is incomplete.
Therefore the protocol is also of repeat-until-success type \cite{Lim2005RepeatUntilSuccess}
in the sense that it can be repeated until the desired degree of parity
projection is reached. We proved these claims analytically for the
ideal case and investigated effects of some of the practical limitations
numerically.

Except for requiring high-quality photodetectors, our protocol is
experimentally implementable in cQED with minimalistic resources.
In particular, the protocol places few requirements on the qubits
and their control lines as it requires only time-independent and dispersive
qubit--resonator coupling.  This is promising for scalability and
calls for a future extension of the protocol
to many-qubit scenarios.
Such an extension would further reduce
the overhead of measuring non-local multi-qubit parity checks for
the purposes of quantum error correction.
\begin{acknowledgments}
We thank Anton Frisk Kockum and Göran Johansson for their help in
comparing our results to Ref.~\cite{FriskKockum2012Undoing}. We
acknowledge financial support from the Emil Aaltonen Foundation, the European
Research Council under Grant 278117 (``SINGLEOUT''), the Academy of
Finland under Grants 135794, 272806, 286215, and 251748 (``COMP''), and
the European Metrology Research Programme (``EXL03 MICROPHOTON'').
The EMRP is jointly
funded by the EMRP participating countries within EURAMET and the
European Union.
\end{acknowledgments}
\appendix

\section{Detector inefficiency and qubit relaxation\label{app:photon-loss}}

\begin{figure}
\includegraphics{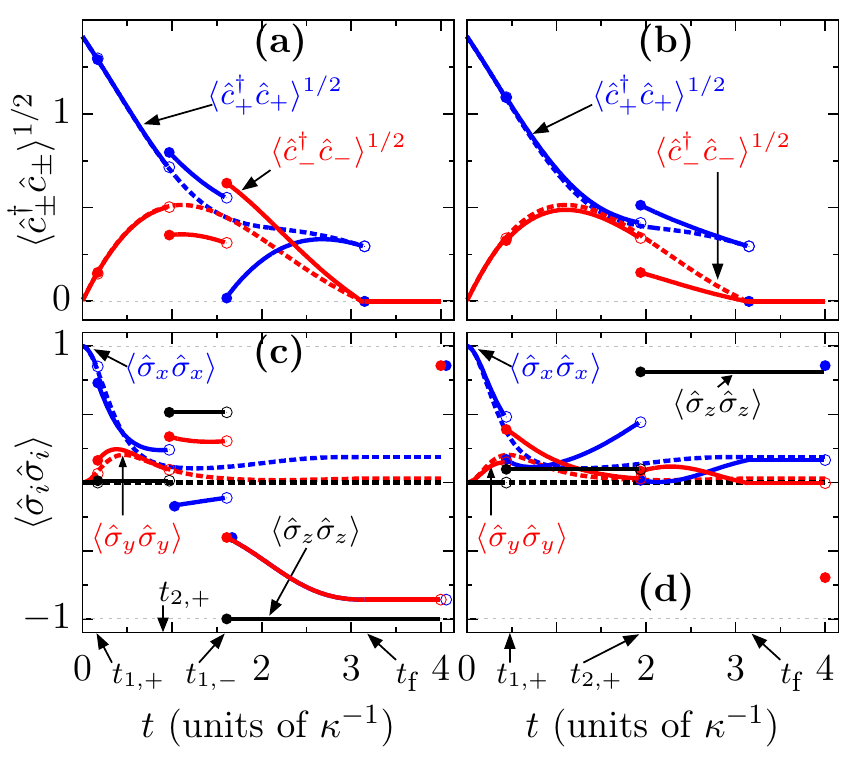}
\caption{(Color online) Trajectories with imperfect detection ($\eta=0.9$
for both detectors), which leads to imperfect reversal of the measurement-induced
dephasing. See Fig.~\ref{fig:trajectories} for
the values of other simulation parameters and
an explanation of
the symbols.\label{fig:inefficient-detection}}
\end{figure}

\begin{figure}
\includegraphics{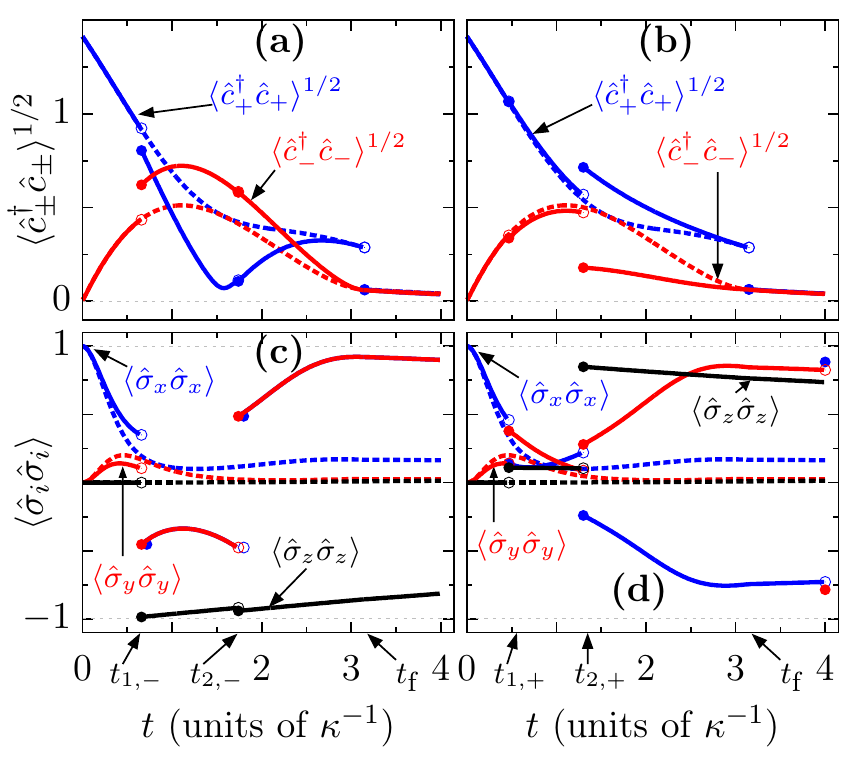}
\caption{(Color online) Trajectories in the presence of qubit relaxation at
a rate $0.1\times\kappa/\pi$. See Fig.~\ref{fig:trajectories} for
the values of other simulation parameters and
an explanation of the symbols.\label{fig:qubit-relaxation}}
\end{figure}

Here we briefly discuss the effect of imperfect photodetection and
of qubit relaxation. We point out some issues that inevitably arise
in experimental realizations but do not attempt to thoroughly map
the effects of non-idealities. We do this by presenting numerical
solutions to Eq.~(\ref{eq:2q-sme}).

Figure~\ref{fig:inefficient-detection} presents example trajectories
for imperfect detection efficiency ($\eta=0.9$). The possibility
of missing photons leads to a mixed state and prevents fully reversing
the measurement-induced dephasing, i.e., $\left\langle \hat{\sigma}_{x,1}\hat{\sigma}_{x,2}\right\rangle <1$
at $t=t_{\textrm{f}}$. However, the unconditioned master equation
{[}Eq.~(\ref{eq:2q-me}){]} is unchanged and the parity of the initial
state is correctly measured, as long as sufficiently many photons
are detected.

Figure~\ref{fig:qubit-relaxation} shows trajectories for a non-zero
qubit relaxation rate. Specifically, we add $\mathcal{D}\left[\sqrt{\mbox{\ensuremath{\gamma}}}\hat{\sigma}_{-,j}\right]\hat{\rho}_{I}$
for each qubit $j$ to the right-hand side of Eq.~(\ref{eq:2q-sme}).
We choose $\gamma=0.1\times\kappa/\pi$. As in the case of inefficient
detection, the qubit relaxation prevents perfect reversal of the measurement-induced
dephasing. In addition, qubit decay leads to a mixed resonator state
even at $t_{\textrm{f}}$. This implies that the displacements in
step \ref{enu:disp-end} of the protocol cannot restore the resonators
to vacuum perfectly, i.e., $\langle \hat{c}_{\pm}^{\dagger}\hat{c}_{\pm}\rangle >0$
even after $t=t_{\textrm{f}}$. Visually the most striking phenomenon
in Fig.~\ref{fig:qubit-relaxation} is the non-conservation of parity
but it occurs even without performing the measurement, i.e., even
if we were to choose $\alpha=0$.

\bibliography{remote_parity_measurement}

\begin{thebibliography}{57}%
\makeatletter
\providecommand \@ifxundefined [1]{%
 \@ifx{#1\undefined}
}%
\providecommand \@ifnum [1]{%
 \ifnum #1\expandafter \@firstoftwo
 \else \expandafter \@secondoftwo
 \fi
}%
\providecommand \@ifx [1]{%
 \ifx #1\expandafter \@firstoftwo
 \else \expandafter \@secondoftwo
 \fi
}%
\providecommand \natexlab [1]{#1}%
\providecommand \enquote  [1]{``#1''}%
\providecommand \bibnamefont  [1]{#1}%
\providecommand \bibfnamefont [1]{#1}%
\providecommand \citenamefont [1]{#1}%
\providecommand \href@noop [0]{\@secondoftwo}%
\providecommand \href [0]{\begingroup \@sanitize@url \@href}%
\providecommand \@href[1]{\@@startlink{#1}\@@href}%
\providecommand \@@href[1]{\endgroup#1\@@endlink}%
\providecommand \@sanitize@url [0]{\catcode `\\12\catcode `\$12\catcode
  `\&12\catcode `\#12\catcode `\^12\catcode `\_12\catcode `\%12\relax}%
\providecommand \@@startlink[1]{}%
\providecommand \@@endlink[0]{}%
\providecommand \url  [0]{\begingroup\@sanitize@url \@url }%
\providecommand \@url [1]{\endgroup\@href {#1}{\urlprefix }}%
\providecommand \urlprefix  [0]{URL }%
\providecommand \Eprint [0]{\href }%
\providecommand \doibase [0]{http://dx.doi.org/}%
\providecommand \selectlanguage [0]{\@gobble}%
\providecommand \bibinfo  [0]{\@secondoftwo}%
\providecommand \bibfield  [0]{\@secondoftwo}%
\providecommand \translation [1]{[#1]}%
\providecommand \BibitemOpen [0]{}%
\providecommand \bibitemStop [0]{}%
\providecommand \bibitemNoStop [0]{.\EOS\space}%
\providecommand \EOS [0]{\spacefactor3000\relax}%
\providecommand \BibitemShut  [1]{\csname bibitem#1\endcsname}%
\let\auto@bib@innerbib\@empty
%</preamble>
\bibitem [{\citenamefont {Cirac}\ and\ \citenamefont
  {Zoller}(1995)}]{Cirac1995Quantum}%
  \BibitemOpen
  \bibfield  {author} {\bibinfo {author} {\bibfnamefont {J.~I.}\ \bibnamefont
  {Cirac}}\ and\ \bibinfo {author} {\bibfnamefont {P.}~\bibnamefont {Zoller}},\
  }\href {\doibase 10.1103/physrevlett.74.4091} {\bibfield  {journal} {\bibinfo
   {journal} {Phys. Rev. Lett.}\ }\textbf {\bibinfo {volume} {74}},\ \bibinfo
  {pages} {4091} (\bibinfo {year} {1995})}\BibitemShut {NoStop}%
\bibitem [{\citenamefont {Cabrillo}\ \emph {et~al.}(1999)\citenamefont
  {Cabrillo}, \citenamefont {Cirac}, \citenamefont
  {Garc\'{\i}a-Fern\'{a}ndez},\ and\ \citenamefont
  {Zoller}}]{Cabrillo1999Creation}%
  \BibitemOpen
  \bibfield  {author} {\bibinfo {author} {\bibfnamefont {C.}~\bibnamefont
  {Cabrillo}}, \bibinfo {author} {\bibfnamefont {J.~I.}\ \bibnamefont {Cirac}},
  \bibinfo {author} {\bibfnamefont {P.}~\bibnamefont
  {Garc\'{\i}a-Fern\'{a}ndez}}, \ and\ \bibinfo {author} {\bibfnamefont
  {P.}~\bibnamefont {Zoller}},\ }\href {\doibase 10.1103/physreva.59.1025}
  {\bibfield  {journal} {\bibinfo  {journal} {Phys. Rev. A}\ }\textbf {\bibinfo
  {volume} {59}},\ \bibinfo {pages} {1025} (\bibinfo {year}
  {1999})}\BibitemShut {NoStop}%
\bibitem [{\citenamefont {Bose}\ \emph {et~al.}(1999)\citenamefont {Bose},
  \citenamefont {Knight}, \citenamefont {Plenio},\ and\ \citenamefont
  {Vedral}}]{Bose1999Proposal}%
  \BibitemOpen
  \bibfield  {author} {\bibinfo {author} {\bibfnamefont {S.}~\bibnamefont
  {Bose}}, \bibinfo {author} {\bibfnamefont {P.~L.}\ \bibnamefont {Knight}},
  \bibinfo {author} {\bibfnamefont {M.~B.}\ \bibnamefont {Plenio}}, \ and\
  \bibinfo {author} {\bibfnamefont {V.}~\bibnamefont {Vedral}},\ }\href
  {\doibase 10.1103/physrevlett.83.5158} {\bibfield  {journal} {\bibinfo
  {journal} {Phys. Rev. Lett.}\ }\textbf {\bibinfo {volume} {83}},\ \bibinfo
  {pages} {5158} (\bibinfo {year} {1999})}\BibitemShut {NoStop}%
\bibitem [{\citenamefont {Duan}\ and\ \citenamefont
  {Kimble}(2003)}]{Duan2003Efficient}%
  \BibitemOpen
  \bibfield  {author} {\bibinfo {author} {\bibfnamefont {L.~M.}\ \bibnamefont
  {Duan}}\ and\ \bibinfo {author} {\bibfnamefont {H.~J.}\ \bibnamefont
  {Kimble}},\ }\href {\doibase 10.1103/physrevlett.90.253601} {\bibfield
  {journal} {\bibinfo  {journal} {Phys. Rev. Lett.}\ }\textbf {\bibinfo
  {volume} {90}},\ \bibinfo {pages} {253601} (\bibinfo {year}
  {2003})}\BibitemShut {NoStop}%
\bibitem [{\citenamefont {Browne}\ \emph {et~al.}(2003)\citenamefont {Browne},
  \citenamefont {Plenio},\ and\ \citenamefont {Huelga}}]{Browne2003Robust}%
  \BibitemOpen
  \bibfield  {author} {\bibinfo {author} {\bibfnamefont {D.~E.}\ \bibnamefont
  {Browne}}, \bibinfo {author} {\bibfnamefont {M.~B.}\ \bibnamefont {Plenio}},
  \ and\ \bibinfo {author} {\bibfnamefont {S.~F.}\ \bibnamefont {Huelga}},\
  }\href {\doibase 10.1103/physrevlett.91.067901} {\bibfield  {journal}
  {\bibinfo  {journal} {Phys. Rev. Lett.}\ }\textbf {\bibinfo {volume} {91}},\
  \bibinfo {pages} {067901} (\bibinfo {year} {2003})}\BibitemShut {NoStop}%
\bibitem [{\citenamefont {Barrett}\ and\ \citenamefont
  {Kok}(2005)}]{Barrett2005Efficient}%
  \BibitemOpen
  \bibfield  {author} {\bibinfo {author} {\bibfnamefont {S.~D.}\ \bibnamefont
  {Barrett}}\ and\ \bibinfo {author} {\bibfnamefont {P.}~\bibnamefont {Kok}},\
  }\href {\doibase 10.1103/physreva.71.060310} {\bibfield  {journal} {\bibinfo
  {journal} {Phys. Rev. A}\ }\textbf {\bibinfo {volume} {71}},\ \bibinfo
  {pages} {060310} (\bibinfo {year} {2005})}\BibitemShut {NoStop}%
\bibitem [{\citenamefont {Lim}\ \emph {et~al.}(2005)\citenamefont {Lim},
  \citenamefont {Beige},\ and\ \citenamefont
  {Kwek}}]{Lim2005RepeatUntilSuccess}%
  \BibitemOpen
  \bibfield  {author} {\bibinfo {author} {\bibfnamefont {Y.~L.}\ \bibnamefont
  {Lim}}, \bibinfo {author} {\bibfnamefont {A.}~\bibnamefont {Beige}}, \ and\
  \bibinfo {author} {\bibfnamefont {L.~C.}\ \bibnamefont {Kwek}},\ }\href
  {\doibase 10.1103/physrevlett.95.030505} {\bibfield  {journal} {\bibinfo
  {journal} {Phys. Rev. Lett.}\ }\textbf {\bibinfo {volume} {95}},\ \bibinfo
  {pages} {030505} (\bibinfo {year} {2005})}\BibitemShut {NoStop}%
\bibitem [{\citenamefont {Ladd}\ \emph {et~al.}(2006)\citenamefont {Ladd},
  \citenamefont {van Loock}, \citenamefont {Nemoto}, \citenamefont {Munro},\
  and\ \citenamefont {Yamamoto}}]{Ladd2006Hybrid}%
  \BibitemOpen
  \bibfield  {author} {\bibinfo {author} {\bibfnamefont {T.~D.}\ \bibnamefont
  {Ladd}}, \bibinfo {author} {\bibfnamefont {P.}~\bibnamefont {van Loock}},
  \bibinfo {author} {\bibfnamefont {K.}~\bibnamefont {Nemoto}}, \bibinfo
  {author} {\bibfnamefont {W.~J.}\ \bibnamefont {Munro}}, \ and\ \bibinfo
  {author} {\bibfnamefont {Y.}~\bibnamefont {Yamamoto}},\ }\href {\doibase
  10.1088/1367-2630/8/9/184} {\bibfield  {journal} {\bibinfo  {journal} {New J.
  Phys.}\ }\textbf {\bibinfo {volume} {8}},\ \bibinfo {pages} {184} (\bibinfo
  {year} {2006})}\BibitemShut {NoStop}%
\bibitem [{\citenamefont {Busch}\ \emph {et~al.}(2008)\citenamefont {Busch},
  \citenamefont {Kyoseva}, \citenamefont {Trupke},\ and\ \citenamefont
  {Beige}}]{Busch2008Entangling}%
  \BibitemOpen
  \bibfield  {author} {\bibinfo {author} {\bibfnamefont {J.}~\bibnamefont
  {Busch}}, \bibinfo {author} {\bibfnamefont {E.~S.}\ \bibnamefont {Kyoseva}},
  \bibinfo {author} {\bibfnamefont {M.}~\bibnamefont {Trupke}}, \ and\ \bibinfo
  {author} {\bibfnamefont {A.}~\bibnamefont {Beige}},\ }\href {\doibase
  10.1103/physreva.78.040301} {\bibfield  {journal} {\bibinfo  {journal} {Phys.
  Rev. A}\ }\textbf {\bibinfo {volume} {78}},\ \bibinfo {pages} {040301}
  (\bibinfo {year} {2008})}\BibitemShut {NoStop}%
\bibitem [{\citenamefont {Matsuzaki}\ \emph {et~al.}(2011)\citenamefont
  {Matsuzaki}, \citenamefont {Benjamin},\ and\ \citenamefont
  {Fitzsimons}}]{Matsuzaki2011Entangling}%
  \BibitemOpen
  \bibfield  {author} {\bibinfo {author} {\bibfnamefont {Y.}~\bibnamefont
  {Matsuzaki}}, \bibinfo {author} {\bibfnamefont {S.~C.}\ \bibnamefont
  {Benjamin}}, \ and\ \bibinfo {author} {\bibfnamefont {J.}~\bibnamefont
  {Fitzsimons}},\ }\href {\doibase 10.1103/physreva.83.060303} {\bibfield
  {journal} {\bibinfo  {journal} {Phys. Rev. A}\ }\textbf {\bibinfo {volume}
  {83}},\ \bibinfo {pages} {060303} (\bibinfo {year} {2011})}\BibitemShut
  {NoStop}%
\bibitem [{\citenamefont {Azuma}\ \emph {et~al.}(2012)\citenamefont {Azuma},
  \citenamefont {Takeda}, \citenamefont {Koashi},\ and\ \citenamefont
  {Imoto}}]{Azuma2012Quantum}%
  \BibitemOpen
  \bibfield  {author} {\bibinfo {author} {\bibfnamefont {K.}~\bibnamefont
  {Azuma}}, \bibinfo {author} {\bibfnamefont {H.}~\bibnamefont {Takeda}},
  \bibinfo {author} {\bibfnamefont {M.}~\bibnamefont {Koashi}}, \ and\ \bibinfo
  {author} {\bibfnamefont {N.}~\bibnamefont {Imoto}},\ }\href {\doibase
  10.1103/physreva.85.062309} {\bibfield  {journal} {\bibinfo  {journal} {Phys.
  Rev. A}\ }\textbf {\bibinfo {volume} {85}},\ \bibinfo {pages} {062309}
  (\bibinfo {year} {2012})}\BibitemShut {NoStop}%
\bibitem [{\citenamefont {Bernien}\ \emph {et~al.}(2013)\citenamefont
  {Bernien}, \citenamefont {Hensen}, \citenamefont {Pfaff}, \citenamefont
  {Koolstra}, \citenamefont {Blok}, \citenamefont {Robledo}, \citenamefont
  {Taminiau}, \citenamefont {Markham}, \citenamefont {Twitchen}, \citenamefont
  {Childress},\ and\ \citenamefont {Hanson}}]{Bernien2013Heralded}%
  \BibitemOpen
  \bibfield  {author} {\bibinfo {author} {\bibfnamefont {H.}~\bibnamefont
  {Bernien}}, \bibinfo {author} {\bibfnamefont {B.}~\bibnamefont {Hensen}},
  \bibinfo {author} {\bibfnamefont {W.}~\bibnamefont {Pfaff}}, \bibinfo
  {author} {\bibfnamefont {G.}~\bibnamefont {Koolstra}}, \bibinfo {author}
  {\bibfnamefont {M.~S.}\ \bibnamefont {Blok}}, \bibinfo {author}
  {\bibfnamefont {L.}~\bibnamefont {Robledo}}, \bibinfo {author} {\bibfnamefont
  {T.~H.}\ \bibnamefont {Taminiau}}, \bibinfo {author} {\bibfnamefont
  {M.}~\bibnamefont {Markham}}, \bibinfo {author} {\bibfnamefont {D.~J.}\
  \bibnamefont {Twitchen}}, \bibinfo {author} {\bibfnamefont {L.}~\bibnamefont
  {Childress}}, \ and\ \bibinfo {author} {\bibfnamefont {R.}~\bibnamefont
  {Hanson}},\ }\href {\doibase 10.1038/nature12016} {\bibfield  {journal}
  {\bibinfo  {journal} {Nature (London)}\ }\textbf {\bibinfo {volume} {497}},\
  \bibinfo {pages} {86} (\bibinfo {year} {2013})}\BibitemShut {NoStop}%
\bibitem [{\citenamefont {Bruschi}\ \emph {et~al.}(2014)\citenamefont
  {Bruschi}, \citenamefont {Barlow}, \citenamefont {Razavi},\ and\
  \citenamefont {Beige}}]{Bruschi2014Repeatuntilsuccess}%
  \BibitemOpen
  \bibfield  {author} {\bibinfo {author} {\bibfnamefont {D.~E.}\ \bibnamefont
  {Bruschi}}, \bibinfo {author} {\bibfnamefont {T.~M.}\ \bibnamefont {Barlow}},
  \bibinfo {author} {\bibfnamefont {M.}~\bibnamefont {Razavi}}, \ and\ \bibinfo
  {author} {\bibfnamefont {A.}~\bibnamefont {Beige}},\ }\href {\doibase
  10.1103/physreva.90.032306} {\bibfield  {journal} {\bibinfo  {journal} {Phys.
  Rev. A}\ }\textbf {\bibinfo {volume} {90}},\ \bibinfo {pages} {032306}
  (\bibinfo {year} {2014})}\BibitemShut {NoStop}%
\bibitem [{\citenamefont {Briegel}\ \emph {et~al.}(1998)\citenamefont
  {Briegel}, \citenamefont {D\"{u}r}, \citenamefont {Cirac},\ and\
  \citenamefont {Zoller}}]{Briegel1998Quantum}%
  \BibitemOpen
  \bibfield  {author} {\bibinfo {author} {\bibfnamefont {H.~J.}\ \bibnamefont
  {Briegel}}, \bibinfo {author} {\bibfnamefont {W.}~\bibnamefont {D\"{u}r}},
  \bibinfo {author} {\bibfnamefont {J.~I.}\ \bibnamefont {Cirac}}, \ and\
  \bibinfo {author} {\bibfnamefont {P.}~\bibnamefont {Zoller}},\ }\href
  {\doibase 10.1103/physrevlett.81.5932} {\bibfield  {journal} {\bibinfo
  {journal} {Phys. Rev. Lett.}\ }\textbf {\bibinfo {volume} {81}},\ \bibinfo
  {pages} {5932} (\bibinfo {year} {1998})}\BibitemShut {NoStop}%
\bibitem [{\citenamefont {Kok}\ \emph {et~al.}(2007)\citenamefont {Kok},
  \citenamefont {Nemoto}, \citenamefont {Ralph}, \citenamefont {Dowling},\ and\
  \citenamefont {Milburn}}]{Kok2007Linear}%
  \BibitemOpen
  \bibfield  {author} {\bibinfo {author} {\bibfnamefont {P.}~\bibnamefont
  {Kok}}, \bibinfo {author} {\bibfnamefont {K.}~\bibnamefont {Nemoto}},
  \bibinfo {author} {\bibfnamefont {T.~C.}\ \bibnamefont {Ralph}}, \bibinfo
  {author} {\bibfnamefont {J.~P.}\ \bibnamefont {Dowling}}, \ and\ \bibinfo
  {author} {\bibfnamefont {G.~J.}\ \bibnamefont {Milburn}},\ }\href {\doibase
  10.1103/revmodphys.79.135} {\bibfield  {journal} {\bibinfo  {journal} {Rev.
  Mod. Phys.}\ }\textbf {\bibinfo {volume} {79}},\ \bibinfo {pages} {135}
  (\bibinfo {year} {2007})}\BibitemShut {NoStop}%
\bibitem [{\citenamefont {Terhal}(2015)}]{Terhal2015Quantum}%
  \BibitemOpen
  \bibfield  {author} {\bibinfo {author} {\bibfnamefont {B.~M.}\ \bibnamefont
  {Terhal}},\ }\href {\doibase 10.1103/revmodphys.87.307} {\bibfield  {journal}
  {\bibinfo  {journal} {Rev. Mod. Phys.}\ }\textbf {\bibinfo {volume} {87}},\
  \bibinfo {pages} {307} (\bibinfo {year} {2015})}\BibitemShut {NoStop}%
\bibitem [{\citenamefont {Bravyi}\ \emph {et~al.}(2010)\citenamefont {Bravyi},
  \citenamefont {Poulin},\ and\ \citenamefont {Terhal}}]{Bravyi2010Tradeoffs}%
  \BibitemOpen
  \bibfield  {author} {\bibinfo {author} {\bibfnamefont {S.}~\bibnamefont
  {Bravyi}}, \bibinfo {author} {\bibfnamefont {D.}~\bibnamefont {Poulin}}, \
  and\ \bibinfo {author} {\bibfnamefont {B.}~\bibnamefont {Terhal}},\ }\href
  {\doibase 10.1103/physrevlett.104.050503} {\bibfield  {journal} {\bibinfo
  {journal} {Phys. Rev. Lett.}\ }\textbf {\bibinfo {volume} {104}},\ \bibinfo
  {pages} {050503} (\bibinfo {year} {2010})}\BibitemShut {NoStop}%
\bibitem [{\citenamefont {Blais}\ \emph {et~al.}(2004)\citenamefont {Blais},
  \citenamefont {Huang}, \citenamefont {Wallraff}, \citenamefont {Girvin},\
  and\ \citenamefont {Schoelkopf}}]{Blais04}%
  \BibitemOpen
  \bibfield  {author} {\bibinfo {author} {\bibfnamefont {A.}~\bibnamefont
  {Blais}}, \bibinfo {author} {\bibfnamefont {R.-S.}\ \bibnamefont {Huang}},
  \bibinfo {author} {\bibfnamefont {A.}~\bibnamefont {Wallraff}}, \bibinfo
  {author} {\bibfnamefont {S.~M.}\ \bibnamefont {Girvin}}, \ and\ \bibinfo
  {author} {\bibfnamefont {R.~J.}\ \bibnamefont {Schoelkopf}},\ }\href
  {\doibase 10.1103/physreva.69.062320} {\bibfield  {journal} {\bibinfo
  {journal} {Phys. Rev. A}\ }\textbf {\bibinfo {volume} {69}},\ \bibinfo
  {pages} {062320} (\bibinfo {year} {2004})}\BibitemShut {NoStop}%
\bibitem [{\citenamefont {Wallraff}\ \emph {et~al.}(2004)\citenamefont
  {Wallraff}, \citenamefont {Schuster}, \citenamefont {Blais}, \citenamefont
  {Frunzio}, \citenamefont {Huang}, \citenamefont {Majer}, \citenamefont
  {Kumar}, \citenamefont {Girvin},\ and\ \citenamefont
  {Schoelkopf}}]{Wallraff2004Strong}%
  \BibitemOpen
  \bibfield  {author} {\bibinfo {author} {\bibfnamefont {A.}~\bibnamefont
  {Wallraff}}, \bibinfo {author} {\bibfnamefont {D.~I.}\ \bibnamefont
  {Schuster}}, \bibinfo {author} {\bibfnamefont {A.}~\bibnamefont {Blais}},
  \bibinfo {author} {\bibfnamefont {L.}~\bibnamefont {Frunzio}}, \bibinfo
  {author} {\bibfnamefont {R.-S.}\ \bibnamefont {Huang}}, \bibinfo {author}
  {\bibfnamefont {J.}~\bibnamefont {Majer}}, \bibinfo {author} {\bibfnamefont
  {S.}~\bibnamefont {Kumar}}, \bibinfo {author} {\bibfnamefont {S.~M.}\
  \bibnamefont {Girvin}}, \ and\ \bibinfo {author} {\bibfnamefont {R.~J.}\
  \bibnamefont {Schoelkopf}},\ }\href {\doibase 10.1038/nature02851} {\bibfield
   {journal} {\bibinfo  {journal} {Nature (London)}\ }\textbf {\bibinfo
  {volume} {431}},\ \bibinfo {pages} {162} (\bibinfo {year}
  {2004})}\BibitemShut {NoStop}%
\bibitem [{\citenamefont {Devoret}\ and\ \citenamefont
  {Schoelkopf}(2013)}]{Devoret2013Superconducting}%
  \BibitemOpen
  \bibfield  {author} {\bibinfo {author} {\bibfnamefont {M.~H.}\ \bibnamefont
  {Devoret}}\ and\ \bibinfo {author} {\bibfnamefont {R.~J.}\ \bibnamefont
  {Schoelkopf}},\ }\href {\doibase 10.1126/science.1231930} {\bibfield
  {journal} {\bibinfo  {journal} {Science}\ }\textbf {\bibinfo {volume}
  {339}},\ \bibinfo {pages} {1169} (\bibinfo {year} {2013})}\BibitemShut
  {NoStop}%
\bibitem [{\citenamefont {Kelly}\ \emph {et~al.}(2015)\citenamefont {Kelly},
  \citenamefont {Barends}, \citenamefont {Fowler}, \citenamefont {Megrant},
  \citenamefont {Jeffrey}, \citenamefont {White}, \citenamefont {Sank},
  \citenamefont {Mutus}, \citenamefont {Campbell}, \citenamefont {Chen},
  \citenamefont {Chen}, \citenamefont {Chiaro}, \citenamefont {Dunsworth},
  \citenamefont {Hoi}, \citenamefont {Neill}, \citenamefont {O'Malley},
  \citenamefont {Quintana}, \citenamefont {Roushan}, \citenamefont
  {Vainsencher}, \citenamefont {Wenner}, \citenamefont {Cleland},\ and\
  \citenamefont {Martinis}}]{Kelly2015State}%
  \BibitemOpen
  \bibfield  {author} {\bibinfo {author} {\bibfnamefont {J.}~\bibnamefont
  {Kelly}}, \bibinfo {author} {\bibfnamefont {R.}~\bibnamefont {Barends}},
  \bibinfo {author} {\bibfnamefont {A.~G.}\ \bibnamefont {Fowler}}, \bibinfo
  {author} {\bibfnamefont {A.}~\bibnamefont {Megrant}}, \bibinfo {author}
  {\bibfnamefont {E.}~\bibnamefont {Jeffrey}}, \bibinfo {author} {\bibfnamefont
  {T.~C.}\ \bibnamefont {White}}, \bibinfo {author} {\bibfnamefont
  {D.}~\bibnamefont {Sank}}, \bibinfo {author} {\bibfnamefont {J.~Y.}\
  \bibnamefont {Mutus}}, \bibinfo {author} {\bibfnamefont {B.}~\bibnamefont
  {Campbell}}, \bibinfo {author} {\bibfnamefont {Y.}~\bibnamefont {Chen}},
  \bibinfo {author} {\bibfnamefont {Z.}~\bibnamefont {Chen}}, \bibinfo {author}
  {\bibfnamefont {B.}~\bibnamefont {Chiaro}}, \bibinfo {author} {\bibfnamefont
  {A.}~\bibnamefont {Dunsworth}}, \bibinfo {author} {\bibfnamefont {I.~C.}\
  \bibnamefont {Hoi}}, \bibinfo {author} {\bibfnamefont {C.}~\bibnamefont
  {Neill}}, \bibinfo {author} {\bibfnamefont {P.~J.~J.}\ \bibnamefont
  {O'Malley}}, \bibinfo {author} {\bibfnamefont {C.}~\bibnamefont {Quintana}},
  \bibinfo {author} {\bibfnamefont {P.}~\bibnamefont {Roushan}}, \bibinfo
  {author} {\bibfnamefont {A.}~\bibnamefont {Vainsencher}}, \bibinfo {author}
  {\bibfnamefont {J.}~\bibnamefont {Wenner}}, \bibinfo {author} {\bibfnamefont
  {A.~N.}\ \bibnamefont {Cleland}}, \ and\ \bibinfo {author} {\bibfnamefont
  {J.~M.}\ \bibnamefont {Martinis}},\ }\href {\doibase 10.1038/nature14270}
  {\bibfield  {journal} {\bibinfo  {journal} {Nature (London)}\ }\textbf
  {\bibinfo {volume} {519}},\ \bibinfo {pages} {66} (\bibinfo {year}
  {2015})}\BibitemShut {NoStop}%
\bibitem [{\citenamefont {Romero}\ \emph {et~al.}(2009)\citenamefont {Romero},
  \citenamefont {Garc\'{\i}a-Ripoll},\ and\ \citenamefont
  {Solano}}]{Romero2009Microwave}%
  \BibitemOpen
  \bibfield  {author} {\bibinfo {author} {\bibfnamefont {G.}~\bibnamefont
  {Romero}}, \bibinfo {author} {\bibfnamefont {J.~J.}\ \bibnamefont
  {Garc\'{\i}a-Ripoll}}, \ and\ \bibinfo {author} {\bibfnamefont
  {E.}~\bibnamefont {Solano}},\ }\href {\doibase
  10.1103/physrevlett.102.173602} {\bibfield  {journal} {\bibinfo  {journal}
  {Phys. Rev. Lett.}\ }\textbf {\bibinfo {volume} {102}},\ \bibinfo {pages}
  {173602} (\bibinfo {year} {2009})}\BibitemShut {NoStop}%
\bibitem [{\citenamefont {Chen}\ \emph {et~al.}(2011)\citenamefont {Chen},
  \citenamefont {Hover}, \citenamefont {Sendelbach}, \citenamefont {Maurer},
  \citenamefont {Merkel}, \citenamefont {Pritchett}, \citenamefont {Wilhelm},\
  and\ \citenamefont {McDermott}}]{Chen2011Microwave}%
  \BibitemOpen
  \bibfield  {author} {\bibinfo {author} {\bibfnamefont {Y.~F.}\ \bibnamefont
  {Chen}}, \bibinfo {author} {\bibfnamefont {D.}~\bibnamefont {Hover}},
  \bibinfo {author} {\bibfnamefont {S.}~\bibnamefont {Sendelbach}}, \bibinfo
  {author} {\bibfnamefont {L.}~\bibnamefont {Maurer}}, \bibinfo {author}
  {\bibfnamefont {S.~T.}\ \bibnamefont {Merkel}}, \bibinfo {author}
  {\bibfnamefont {E.~J.}\ \bibnamefont {Pritchett}}, \bibinfo {author}
  {\bibfnamefont {F.~K.}\ \bibnamefont {Wilhelm}}, \ and\ \bibinfo {author}
  {\bibfnamefont {R.}~\bibnamefont {McDermott}},\ }\href {\doibase
  10.1103/physrevlett.107.217401} {\bibfield  {journal} {\bibinfo  {journal}
  {Phys. Rev. Lett.}\ }\textbf {\bibinfo {volume} {107}},\ \bibinfo {pages}
  {217401} (\bibinfo {year} {2011})}\BibitemShut {NoStop}%
\bibitem [{\citenamefont {Peropadre}\ \emph {et~al.}(2011)\citenamefont
  {Peropadre}, \citenamefont {Romero}, \citenamefont {Johansson}, \citenamefont
  {Wilson}, \citenamefont {Solano},\ and\ \citenamefont
  {Garc\'{\i}a-Ripoll}}]{Peropadre2011Approaching}%
  \BibitemOpen
  \bibfield  {author} {\bibinfo {author} {\bibfnamefont {B.}~\bibnamefont
  {Peropadre}}, \bibinfo {author} {\bibfnamefont {G.}~\bibnamefont {Romero}},
  \bibinfo {author} {\bibfnamefont {G.}~\bibnamefont {Johansson}}, \bibinfo
  {author} {\bibfnamefont {C.~M.}\ \bibnamefont {Wilson}}, \bibinfo {author}
  {\bibfnamefont {E.}~\bibnamefont {Solano}}, \ and\ \bibinfo {author}
  {\bibfnamefont {J.~J.}\ \bibnamefont {Garc\'{\i}a-Ripoll}},\ }\href {\doibase
  10.1103/physreva.84.063834} {\bibfield  {journal} {\bibinfo  {journal} {Phys.
  Rev. A}\ }\textbf {\bibinfo {volume} {84}},\ \bibinfo {pages} {063834}
  (\bibinfo {year} {2011})}\BibitemShut {NoStop}%
\bibitem [{\citenamefont {Fan}\ \emph {et~al.}(2013)\citenamefont {Fan},
  \citenamefont {Kockum}, \citenamefont {Combes}, \citenamefont {Johansson},
  \citenamefont {Hoi}, \citenamefont {Wilson}, \citenamefont {Delsing},
  \citenamefont {Milburn},\ and\ \citenamefont {Stace}}]{Fan2013Breakdown}%
  \BibitemOpen
  \bibfield  {author} {\bibinfo {author} {\bibfnamefont {B.}~\bibnamefont
  {Fan}}, \bibinfo {author} {\bibfnamefont {A.~F.}\ \bibnamefont {Kockum}},
  \bibinfo {author} {\bibfnamefont {J.}~\bibnamefont {Combes}}, \bibinfo
  {author} {\bibfnamefont {G.}~\bibnamefont {Johansson}}, \bibinfo {author}
  {\bibfnamefont {I.-C.}\ \bibnamefont {Hoi}}, \bibinfo {author} {\bibfnamefont
  {C.~M.}\ \bibnamefont {Wilson}}, \bibinfo {author} {\bibfnamefont
  {P.}~\bibnamefont {Delsing}}, \bibinfo {author} {\bibfnamefont {G.~J.}\
  \bibnamefont {Milburn}}, \ and\ \bibinfo {author} {\bibfnamefont {T.~M.}\
  \bibnamefont {Stace}},\ }\href {\doibase 10.1103/physrevlett.110.053601}
  {\bibfield  {journal} {\bibinfo  {journal} {Phys. Rev. Lett.}\ }\textbf
  {\bibinfo {volume} {110}},\ \bibinfo {pages} {053601} (\bibinfo {year}
  {2013})}\BibitemShut {NoStop}%
\bibitem [{\citenamefont {Hoi}\ \emph {et~al.}(2013)\citenamefont {Hoi},
  \citenamefont {Kockum}, \citenamefont {Palomaki}, \citenamefont {Stace},
  \citenamefont {Fan}, \citenamefont {Tornberg}, \citenamefont {Sathyamoorthy},
  \citenamefont {Johansson}, \citenamefont {Delsing},\ and\ \citenamefont
  {Wilson}}]{Hoi2013Giant}%
  \BibitemOpen
  \bibfield  {author} {\bibinfo {author} {\bibfnamefont {I.-C.}\ \bibnamefont
  {Hoi}}, \bibinfo {author} {\bibfnamefont {A.~F.}\ \bibnamefont {Kockum}},
  \bibinfo {author} {\bibfnamefont {T.}~\bibnamefont {Palomaki}}, \bibinfo
  {author} {\bibfnamefont {T.~M.}\ \bibnamefont {Stace}}, \bibinfo {author}
  {\bibfnamefont {B.}~\bibnamefont {Fan}}, \bibinfo {author} {\bibfnamefont
  {L.}~\bibnamefont {Tornberg}}, \bibinfo {author} {\bibfnamefont {S.~R.}\
  \bibnamefont {Sathyamoorthy}}, \bibinfo {author} {\bibfnamefont
  {G.}~\bibnamefont {Johansson}}, \bibinfo {author} {\bibfnamefont
  {P.}~\bibnamefont {Delsing}}, \ and\ \bibinfo {author} {\bibfnamefont
  {C.~M.}\ \bibnamefont {Wilson}},\ }\href {\doibase
  10.1103/physrevlett.111.053601} {\bibfield  {journal} {\bibinfo  {journal}
  {Phys. Rev. Lett.}\ }\textbf {\bibinfo {volume} {111}},\ \bibinfo {pages}
  {053601} (\bibinfo {year} {2013})}\BibitemShut {NoStop}%
\bibitem [{\citenamefont {Sathyamoorthy}\ \emph {et~al.}(2014)\citenamefont
  {Sathyamoorthy}, \citenamefont {Tornberg}, \citenamefont {Kockum},
  \citenamefont {Baragiola}, \citenamefont {Combes}, \citenamefont {Wilson},
  \citenamefont {Stace},\ and\ \citenamefont
  {Johansson}}]{Sathyamoorthy2014Quantum}%
  \BibitemOpen
  \bibfield  {author} {\bibinfo {author} {\bibfnamefont {S.~R.}\ \bibnamefont
  {Sathyamoorthy}}, \bibinfo {author} {\bibfnamefont {L.}~\bibnamefont
  {Tornberg}}, \bibinfo {author} {\bibfnamefont {A.~F.}\ \bibnamefont
  {Kockum}}, \bibinfo {author} {\bibfnamefont {B.~Q.}\ \bibnamefont
  {Baragiola}}, \bibinfo {author} {\bibfnamefont {J.}~\bibnamefont {Combes}},
  \bibinfo {author} {\bibfnamefont {C.~M.}\ \bibnamefont {Wilson}}, \bibinfo
  {author} {\bibfnamefont {T.~M.}\ \bibnamefont {Stace}}, \ and\ \bibinfo
  {author} {\bibfnamefont {G.}~\bibnamefont {Johansson}},\ }\href {\doibase
  10.1103/physrevlett.112.093601} {\bibfield  {journal} {\bibinfo  {journal}
  {Phys. Rev. Lett.}\ }\textbf {\bibinfo {volume} {112}},\ \bibinfo {pages}
  {093601} (\bibinfo {year} {2014})}\BibitemShut {NoStop}%
\bibitem [{\citenamefont {Fan}\ \emph {et~al.}(2014)\citenamefont {Fan},
  \citenamefont {Johansson}, \citenamefont {Combes}, \citenamefont {Milburn},\
  and\ \citenamefont {Stace}}]{Fan2014Nonabsorbing}%
  \BibitemOpen
  \bibfield  {author} {\bibinfo {author} {\bibfnamefont {B.}~\bibnamefont
  {Fan}}, \bibinfo {author} {\bibfnamefont {G.}~\bibnamefont {Johansson}},
  \bibinfo {author} {\bibfnamefont {J.}~\bibnamefont {Combes}}, \bibinfo
  {author} {\bibfnamefont {G.~J.}\ \bibnamefont {Milburn}}, \ and\ \bibinfo
  {author} {\bibfnamefont {T.~M.}\ \bibnamefont {Stace}},\ }\href {\doibase
  10.1103/physrevb.90.035132} {\bibfield  {journal} {\bibinfo  {journal} {Phys.
  Rev. B}\ }\textbf {\bibinfo {volume} {90}},\ \bibinfo {pages} {035132}
  (\bibinfo {year} {2014})}\BibitemShut {NoStop}%
\bibitem [{\citenamefont {Govenius}\ \emph {et~al.}(2014)\citenamefont
  {Govenius}, \citenamefont {Lake}, \citenamefont {Tan}, \citenamefont
  {Pietil\"{a}}, \citenamefont {Julin}, \citenamefont {Maasilta}, \citenamefont
  {Virtanen},\ and\ \citenamefont {M\"{o}tt\"{o}nen}}]{Govenius2014Microwave}%
  \BibitemOpen
  \bibfield  {author} {\bibinfo {author} {\bibfnamefont {J.}~\bibnamefont
  {Govenius}}, \bibinfo {author} {\bibfnamefont {R.~E.}\ \bibnamefont {Lake}},
  \bibinfo {author} {\bibfnamefont {K.~Y.}\ \bibnamefont {Tan}}, \bibinfo
  {author} {\bibfnamefont {V.}~\bibnamefont {Pietil\"{a}}}, \bibinfo {author}
  {\bibfnamefont {J.~K.}\ \bibnamefont {Julin}}, \bibinfo {author}
  {\bibfnamefont {I.~J.}\ \bibnamefont {Maasilta}}, \bibinfo {author}
  {\bibfnamefont {P.}~\bibnamefont {Virtanen}}, \ and\ \bibinfo {author}
  {\bibfnamefont {M.}~\bibnamefont {M\"{o}tt\"{o}nen}},\ }\href {\doibase
  10.1103/physrevb.90.064505} {\bibfield  {journal} {\bibinfo  {journal} {Phys.
  Rev. B}\ }\textbf {\bibinfo {volume} {90}},\ \bibinfo {pages} {064505}
  (\bibinfo {year} {2014})}\BibitemShut {NoStop}%
\bibitem [{\citenamefont {Gasparinetti}\ \emph {et~al.}(2015)\citenamefont
  {Gasparinetti}, \citenamefont {Viisanen}, \citenamefont {Saira},
  \citenamefont {Faivre}, \citenamefont {Arzeo}, \citenamefont {Meschke},\ and\
  \citenamefont {Pekola}}]{Gasparinetti2015Fast}%
  \BibitemOpen
  \bibfield  {author} {\bibinfo {author} {\bibfnamefont {S.}~\bibnamefont
  {Gasparinetti}}, \bibinfo {author} {\bibfnamefont {K.~L.}\ \bibnamefont
  {Viisanen}}, \bibinfo {author} {\bibfnamefont {O.-P.}\ \bibnamefont {Saira}},
  \bibinfo {author} {\bibfnamefont {T.}~\bibnamefont {Faivre}}, \bibinfo
  {author} {\bibfnamefont {M.}~\bibnamefont {Arzeo}}, \bibinfo {author}
  {\bibfnamefont {M.}~\bibnamefont {Meschke}}, \ and\ \bibinfo {author}
  {\bibfnamefont {J.~P.}\ \bibnamefont {Pekola}},\ }\href {\doibase
  10.1103/physrevapplied.3.014007} {\bibfield  {journal} {\bibinfo  {journal}
  {Phys. Rev. Appl.}\ }\textbf {\bibinfo {volume} {3}},\ \bibinfo {pages}
  {014007} (\bibinfo {year} {2015})}\BibitemShut {NoStop}%
\bibitem [{\citenamefont {Koshino}\ \emph {et~al.}(2015)\citenamefont
  {Koshino}, \citenamefont {Inomata}, \citenamefont {Lin}, \citenamefont
  {Nakamura},\ and\ \citenamefont {Yamamoto}}]{Koshino2015Theory}%
  \BibitemOpen
  \bibfield  {author} {\bibinfo {author} {\bibfnamefont {K.}~\bibnamefont
  {Koshino}}, \bibinfo {author} {\bibfnamefont {K.}~\bibnamefont {Inomata}},
  \bibinfo {author} {\bibfnamefont {Z.}~\bibnamefont {Lin}}, \bibinfo {author}
  {\bibfnamefont {Y.}~\bibnamefont {Nakamura}}, \ and\ \bibinfo {author}
  {\bibfnamefont {T.}~\bibnamefont {Yamamoto}},\ }\href {\doibase
  10.1103/physreva.91.043805} {\bibfield  {journal} {\bibinfo  {journal} {Phys.
  Rev. A}\ }\textbf {\bibinfo {volume} {91}},\ \bibinfo {pages} {043805}
  (\bibinfo {year} {2015})}\BibitemShut {NoStop}%
\bibitem [{\citenamefont {Schuster}\ \emph {et~al.}(2007)\citenamefont
  {Schuster}, \citenamefont {Houck}, \citenamefont {Schreier}, \citenamefont
  {Wallraff}, \citenamefont {Gambetta}, \citenamefont {Blais}, \citenamefont
  {Frunzio}, \citenamefont {Majer}, \citenamefont {Johnson}, \citenamefont
  {Devoret}, \citenamefont {Girvin},\ and\ \citenamefont
  {Schoelkopf}}]{Schuster2007Resolving}%
  \BibitemOpen
  \bibfield  {author} {\bibinfo {author} {\bibfnamefont {D.~I.}\ \bibnamefont
  {Schuster}}, \bibinfo {author} {\bibfnamefont {A.~A.}\ \bibnamefont {Houck}},
  \bibinfo {author} {\bibfnamefont {J.~A.}\ \bibnamefont {Schreier}}, \bibinfo
  {author} {\bibfnamefont {A.}~\bibnamefont {Wallraff}}, \bibinfo {author}
  {\bibfnamefont {J.~M.}\ \bibnamefont {Gambetta}}, \bibinfo {author}
  {\bibfnamefont {A.}~\bibnamefont {Blais}}, \bibinfo {author} {\bibfnamefont
  {L.}~\bibnamefont {Frunzio}}, \bibinfo {author} {\bibfnamefont
  {J.}~\bibnamefont {Majer}}, \bibinfo {author} {\bibfnamefont
  {B.}~\bibnamefont {Johnson}}, \bibinfo {author} {\bibfnamefont {M.~H.}\
  \bibnamefont {Devoret}}, \bibinfo {author} {\bibfnamefont {S.~M.}\
  \bibnamefont {Girvin}}, \ and\ \bibinfo {author} {\bibfnamefont {R.~J.}\
  \bibnamefont {Schoelkopf}},\ }\href {\doibase 10.1038/nature05461} {\bibfield
   {journal} {\bibinfo  {journal} {Nature (London)}\ }\textbf {\bibinfo
  {volume} {445}},\ \bibinfo {pages} {515} (\bibinfo {year}
  {2007})}\BibitemShut {NoStop}%
\bibitem [{\citenamefont {Kerckhoff}\ \emph {et~al.}(2009)\citenamefont
  {Kerckhoff}, \citenamefont {Bouten}, \citenamefont {Silberfarb},\ and\
  \citenamefont {Mabuchi}}]{Kerckhoff2009Physical}%
  \BibitemOpen
  \bibfield  {author} {\bibinfo {author} {\bibfnamefont {J.}~\bibnamefont
  {Kerckhoff}}, \bibinfo {author} {\bibfnamefont {L.}~\bibnamefont {Bouten}},
  \bibinfo {author} {\bibfnamefont {A.}~\bibnamefont {Silberfarb}}, \ and\
  \bibinfo {author} {\bibfnamefont {H.}~\bibnamefont {Mabuchi}},\ }\href
  {\doibase 10.1103/physreva.79.024305} {\bibfield  {journal} {\bibinfo
  {journal} {Phys. Rev. A}\ }\textbf {\bibinfo {volume} {79}},\ \bibinfo
  {pages} {024305} (\bibinfo {year} {2009})}\BibitemShut {NoStop}%
\bibitem [{\citenamefont {Roch}\ \emph {et~al.}(2014)\citenamefont {Roch},
  \citenamefont {Schwartz}, \citenamefont {Motzoi}, \citenamefont {Macklin},
  \citenamefont {Vijay}, \citenamefont {Eddins}, \citenamefont {Korotkov},
  \citenamefont {Whaley}, \citenamefont {Sarovar},\ and\ \citenamefont
  {Siddiqi}}]{Roch2014Observation}%
  \BibitemOpen
  \bibfield  {author} {\bibinfo {author} {\bibfnamefont {N.}~\bibnamefont
  {Roch}}, \bibinfo {author} {\bibfnamefont {M.~E.}\ \bibnamefont {Schwartz}},
  \bibinfo {author} {\bibfnamefont {F.}~\bibnamefont {Motzoi}}, \bibinfo
  {author} {\bibfnamefont {C.}~\bibnamefont {Macklin}}, \bibinfo {author}
  {\bibfnamefont {R.}~\bibnamefont {Vijay}}, \bibinfo {author} {\bibfnamefont
  {A.~W.}\ \bibnamefont {Eddins}}, \bibinfo {author} {\bibfnamefont {A.~N.}\
  \bibnamefont {Korotkov}}, \bibinfo {author} {\bibfnamefont {K.~B.}\
  \bibnamefont {Whaley}}, \bibinfo {author} {\bibfnamefont {M.}~\bibnamefont
  {Sarovar}}, \ and\ \bibinfo {author} {\bibfnamefont {I.}~\bibnamefont
  {Siddiqi}},\ }\href {\doibase 10.1103/physrevlett.112.170501} {\bibfield
  {journal} {\bibinfo  {journal} {Phys. Rev. Lett.}\ }\textbf {\bibinfo
  {volume} {112}},\ \bibinfo {pages} {170501} (\bibinfo {year}
  {2014})}\BibitemShut {NoStop}%
\bibitem [{\citenamefont {Schuster}\ \emph {et~al.}(2005)\citenamefont
  {Schuster}, \citenamefont {Wallraff}, \citenamefont {Blais}, \citenamefont
  {Frunzio}, \citenamefont {Huang}, \citenamefont {Majer}, \citenamefont
  {Girvin},\ and\ \citenamefont {Schoelkopf}}]{Schuster2005Ac}%
  \BibitemOpen
  \bibfield  {author} {\bibinfo {author} {\bibfnamefont {D.~I.}\ \bibnamefont
  {Schuster}}, \bibinfo {author} {\bibfnamefont {A.}~\bibnamefont {Wallraff}},
  \bibinfo {author} {\bibfnamefont {A.}~\bibnamefont {Blais}}, \bibinfo
  {author} {\bibfnamefont {L.}~\bibnamefont {Frunzio}}, \bibinfo {author}
  {\bibfnamefont {R.-S.}\ \bibnamefont {Huang}}, \bibinfo {author}
  {\bibfnamefont {J.}~\bibnamefont {Majer}}, \bibinfo {author} {\bibfnamefont
  {S.~M.}\ \bibnamefont {Girvin}}, \ and\ \bibinfo {author} {\bibfnamefont
  {R.~J.}\ \bibnamefont {Schoelkopf}},\ }\href {\doibase
  10.1103/physrevlett.94.123602} {\bibfield  {journal} {\bibinfo  {journal}
  {Phys. Rev. Lett.}\ }\textbf {\bibinfo {volume} {94}},\ \bibinfo {pages}
  {123602} (\bibinfo {year} {2005})}\BibitemShut {NoStop}%
\bibitem [{\citenamefont {Gambetta}\ \emph {et~al.}(2006)\citenamefont
  {Gambetta}, \citenamefont {Blais}, \citenamefont {Schuster}, \citenamefont
  {Wallraff}, \citenamefont {Frunzio}, \citenamefont {Majer}, \citenamefont
  {Devoret}, \citenamefont {Girvin},\ and\ \citenamefont
  {Schoelkopf}}]{Gambetta2006Qubitphoton}%
  \BibitemOpen
  \bibfield  {author} {\bibinfo {author} {\bibfnamefont {J.}~\bibnamefont
  {Gambetta}}, \bibinfo {author} {\bibfnamefont {A.}~\bibnamefont {Blais}},
  \bibinfo {author} {\bibfnamefont {D.~I.}\ \bibnamefont {Schuster}}, \bibinfo
  {author} {\bibfnamefont {A.}~\bibnamefont {Wallraff}}, \bibinfo {author}
  {\bibfnamefont {L.}~\bibnamefont {Frunzio}}, \bibinfo {author} {\bibfnamefont
  {J.}~\bibnamefont {Majer}}, \bibinfo {author} {\bibfnamefont {M.~H.}\
  \bibnamefont {Devoret}}, \bibinfo {author} {\bibfnamefont {S.~M.}\
  \bibnamefont {Girvin}}, \ and\ \bibinfo {author} {\bibfnamefont {R.~J.}\
  \bibnamefont {Schoelkopf}},\ }\href {\doibase 10.1103/physreva.74.042318}
  {\bibfield  {journal} {\bibinfo  {journal} {Phys. Rev. A}\ }\textbf {\bibinfo
  {volume} {74}},\ \bibinfo {pages} {042318} (\bibinfo {year}
  {2006})}\BibitemShut {NoStop}%
\bibitem [{\citenamefont {Gambetta}\ \emph {et~al.}(2008)\citenamefont
  {Gambetta}, \citenamefont {Blais}, \citenamefont {Boissonneault},
  \citenamefont {Houck}, \citenamefont {Schuster},\ and\ \citenamefont
  {Girvin}}]{Gambetta2008Quantum}%
  \BibitemOpen
  \bibfield  {author} {\bibinfo {author} {\bibfnamefont {J.}~\bibnamefont
  {Gambetta}}, \bibinfo {author} {\bibfnamefont {A.}~\bibnamefont {Blais}},
  \bibinfo {author} {\bibfnamefont {M.}~\bibnamefont {Boissonneault}}, \bibinfo
  {author} {\bibfnamefont {A.~A.}\ \bibnamefont {Houck}}, \bibinfo {author}
  {\bibfnamefont {D.~I.}\ \bibnamefont {Schuster}}, \ and\ \bibinfo {author}
  {\bibfnamefont {S.~M.}\ \bibnamefont {Girvin}},\ }\href {\doibase
  10.1103/physreva.77.012112} {\bibfield  {journal} {\bibinfo  {journal} {Phys.
  Rev. A}\ }\textbf {\bibinfo {volume} {77}},\ \bibinfo {pages} {012112}
  (\bibinfo {year} {2008})}\BibitemShut {NoStop}%
\bibitem [{\citenamefont {Lalumi\`{e}re}\ \emph {et~al.}(2010)\citenamefont
  {Lalumi\`{e}re}, \citenamefont {Gambetta},\ and\ \citenamefont
  {Blais}}]{Lalumiere2010Tunable}%
  \BibitemOpen
  \bibfield  {author} {\bibinfo {author} {\bibfnamefont {K.}~\bibnamefont
  {Lalumi\`{e}re}}, \bibinfo {author} {\bibfnamefont {J.~M.}\ \bibnamefont
  {Gambetta}}, \ and\ \bibinfo {author} {\bibfnamefont {A.}~\bibnamefont
  {Blais}},\ }\href {\doibase 10.1103/physreva.81.040301} {\bibfield  {journal}
  {\bibinfo  {journal} {Phys. Rev. A}\ }\textbf {\bibinfo {volume} {81}},\
  \bibinfo {pages} {040301} (\bibinfo {year} {2010})}\BibitemShut {NoStop}%
\bibitem [{\citenamefont {Tornberg}\ and\ \citenamefont
  {Johansson}(2010)}]{Tornberg2010Highfidelity}%
  \BibitemOpen
  \bibfield  {author} {\bibinfo {author} {\bibfnamefont {L.}~\bibnamefont
  {Tornberg}}\ and\ \bibinfo {author} {\bibfnamefont {G.}~\bibnamefont
  {Johansson}},\ }\href {\doibase 10.1103/physreva.82.012329} {\bibfield
  {journal} {\bibinfo  {journal} {Phys. Rev. A}\ }\textbf {\bibinfo {volume}
  {82}},\ \bibinfo {pages} {012329} (\bibinfo {year} {2010})}\BibitemShut
  {NoStop}%
\bibitem [{\citenamefont {Riste}\ \emph {et~al.}(2013)\citenamefont {Riste},
  \citenamefont {Dukalski}, \citenamefont {Watson}, \citenamefont {de~Lange},
  \citenamefont {Tiggelman}, \citenamefont {Blanter}, \citenamefont {Lehnert},
  \citenamefont {Schouten},\ and\ \citenamefont
  {DiCarlo}}]{Riste2013Deterministic}%
  \BibitemOpen
  \bibfield  {author} {\bibinfo {author} {\bibfnamefont {D.}~\bibnamefont
  {Riste}}, \bibinfo {author} {\bibfnamefont {M.}~\bibnamefont {Dukalski}},
  \bibinfo {author} {\bibfnamefont {C.~A.}\ \bibnamefont {Watson}}, \bibinfo
  {author} {\bibfnamefont {G.}~\bibnamefont {de~Lange}}, \bibinfo {author}
  {\bibfnamefont {M.~J.}\ \bibnamefont {Tiggelman}}, \bibinfo {author}
  {\bibfnamefont {Y.}~\bibnamefont {Blanter}}, \bibinfo {author} {\bibfnamefont
  {K.~W.}\ \bibnamefont {Lehnert}}, \bibinfo {author} {\bibfnamefont {R.~N.}\
  \bibnamefont {Schouten}}, \ and\ \bibinfo {author} {\bibfnamefont
  {L.}~\bibnamefont {DiCarlo}},\ }\href {\doibase 10.1038/nature12513}
  {\bibfield  {journal} {\bibinfo  {journal} {Nature (London)}\ }\textbf
  {\bibinfo {volume} {502}},\ \bibinfo {pages} {350} (\bibinfo {year}
  {2013})}\BibitemShut {NoStop}%
\bibitem [{\citenamefont {Korotkov}\ and\ \citenamefont
  {Jordan}(2006)}]{Korotkov2006Undoing}%
  \BibitemOpen
  \bibfield  {author} {\bibinfo {author} {\bibfnamefont {A.~N.}\ \bibnamefont
  {Korotkov}}\ and\ \bibinfo {author} {\bibfnamefont {A.~N.}\ \bibnamefont
  {Jordan}},\ }\href {\doibase 10.1103/physrevlett.97.166805} {\bibfield
  {journal} {\bibinfo  {journal} {Phys. Rev. Lett.}\ }\textbf {\bibinfo
  {volume} {97}},\ \bibinfo {pages} {166805} (\bibinfo {year}
  {2006})}\BibitemShut {NoStop}%
\bibitem [{\citenamefont {Frisk~Kockum}\ \emph {et~al.}(2012)\citenamefont
  {Frisk~Kockum}, \citenamefont {Tornberg},\ and\ \citenamefont
  {Johansson}}]{FriskKockum2012Undoing}%
  \BibitemOpen
  \bibfield  {author} {\bibinfo {author} {\bibfnamefont {A.}~\bibnamefont
  {Frisk~Kockum}}, \bibinfo {author} {\bibfnamefont {L.}~\bibnamefont
  {Tornberg}}, \ and\ \bibinfo {author} {\bibfnamefont {G.}~\bibnamefont
  {Johansson}},\ }\href {\doibase 10.1103/physreva.85.052318} {\bibfield
  {journal} {\bibinfo  {journal} {Phys. Rev. A}\ }\textbf {\bibinfo {volume}
  {85}},\ \bibinfo {pages} {052318} (\bibinfo {year} {2012})}\BibitemShut
  {NoStop}%
\bibitem [{\citenamefont {de~Lange}\ \emph {et~al.}(2014)\citenamefont
  {de~Lange}, \citenamefont {Rist\`{e}}, \citenamefont {Tiggelman},
  \citenamefont {Eichler}, \citenamefont {Tornberg}, \citenamefont {Johansson},
  \citenamefont {Wallraff}, \citenamefont {Schouten},\ and\ \citenamefont
  {DiCarlo}}]{deLange2014Reversing}%
  \BibitemOpen
  \bibfield  {author} {\bibinfo {author} {\bibfnamefont {G.}~\bibnamefont
  {de~Lange}}, \bibinfo {author} {\bibfnamefont {D.}~\bibnamefont {Rist\`{e}}},
  \bibinfo {author} {\bibfnamefont {M.~J.}\ \bibnamefont {Tiggelman}}, \bibinfo
  {author} {\bibfnamefont {C.}~\bibnamefont {Eichler}}, \bibinfo {author}
  {\bibfnamefont {L.}~\bibnamefont {Tornberg}}, \bibinfo {author}
  {\bibfnamefont {G.}~\bibnamefont {Johansson}}, \bibinfo {author}
  {\bibfnamefont {A.}~\bibnamefont {Wallraff}}, \bibinfo {author}
  {\bibfnamefont {R.~N.}\ \bibnamefont {Schouten}}, \ and\ \bibinfo {author}
  {\bibfnamefont {L.}~\bibnamefont {DiCarlo}},\ }\href {\doibase
  10.1103/physrevlett.112.080501} {\bibfield  {journal} {\bibinfo  {journal}
  {Phys. Rev. Lett.}\ }\textbf {\bibinfo {volume} {112}},\ \bibinfo {pages}
  {080501} (\bibinfo {year} {2014})}\BibitemShut {NoStop}%
\bibitem [{\citenamefont {Clerk}\ \emph {et~al.}(2010)\citenamefont {Clerk},
  \citenamefont {Devoret}, \citenamefont {Girvin}, \citenamefont {Marquardt},\
  and\ \citenamefont {Schoelkopf}}]{Clerk2010Introduction}%
  \BibitemOpen
  \bibfield  {author} {\bibinfo {author} {\bibfnamefont {A.~A.}\ \bibnamefont
  {Clerk}}, \bibinfo {author} {\bibfnamefont {M.~H.}\ \bibnamefont {Devoret}},
  \bibinfo {author} {\bibfnamefont {S.~M.}\ \bibnamefont {Girvin}}, \bibinfo
  {author} {\bibfnamefont {F.}~\bibnamefont {Marquardt}}, \ and\ \bibinfo
  {author} {\bibfnamefont {R.~J.}\ \bibnamefont {Schoelkopf}},\ }\href
  {\doibase 10.1103/revmodphys.82.1155} {\bibfield  {journal} {\bibinfo
  {journal} {Rev. Mod. Phys.}\ }\textbf {\bibinfo {volume} {82}},\ \bibinfo
  {pages} {1155} (\bibinfo {year} {2010})}\BibitemShut {NoStop}%
\bibitem [{\citenamefont {Wiseman}\ and\ \citenamefont
  {Milburn}(2010)}]{Wiseman2009Quantum}%
  \BibitemOpen
  \bibfield  {author} {\bibinfo {author} {\bibfnamefont {H.~M.}\ \bibnamefont
  {Wiseman}}\ and\ \bibinfo {author} {\bibfnamefont {G.~J.}\ \bibnamefont
  {Milburn}},\ }\href@noop {} {\emph {\bibinfo {title} {Quantum measurement and
  control}}}\ (\bibinfo  {publisher} {Cambridge University Press},\ \bibinfo
  {year} {2010})\BibitemShut {NoStop}%
\bibitem [{\citenamefont {Hutchison}\ \emph {et~al.}(2009)\citenamefont
  {Hutchison}, \citenamefont {Gambetta}, \citenamefont {Blais},\ and\
  \citenamefont {Wilhelm}}]{Hutchison2009Quantum}%
  \BibitemOpen
  \bibfield  {author} {\bibinfo {author} {\bibfnamefont {C.~L.}\ \bibnamefont
  {Hutchison}}, \bibinfo {author} {\bibfnamefont {J.~M.}\ \bibnamefont
  {Gambetta}}, \bibinfo {author} {\bibfnamefont {A.}~\bibnamefont {Blais}}, \
  and\ \bibinfo {author} {\bibfnamefont {F.~K.}\ \bibnamefont {Wilhelm}},\
  }\href {\doibase 10.1139/p08-140} {\bibfield  {journal} {\bibinfo  {journal}
  {Can. J. Phys.}\ }\textbf {\bibinfo {volume} {87}},\ \bibinfo {pages} {225}
  (\bibinfo {year} {2009})}\BibitemShut {NoStop}%
\bibitem [{\citenamefont {Bishop}\ \emph {et~al.}(2009)\citenamefont {Bishop},
  \citenamefont {Tornberg}, \citenamefont {Price}, \citenamefont {Ginossar},
  \citenamefont {Nunnenkamp}, \citenamefont {Houck}, \citenamefont {Gambetta},
  \citenamefont {Koch}, \citenamefont {Johansson}, \citenamefont {Girvin},\
  and\ \citenamefont {Schoelkopf}}]{Bishop2009Proposal}%
  \BibitemOpen
  \bibfield  {author} {\bibinfo {author} {\bibfnamefont {L.~S.}\ \bibnamefont
  {Bishop}}, \bibinfo {author} {\bibfnamefont {L.}~\bibnamefont {Tornberg}},
  \bibinfo {author} {\bibfnamefont {D.}~\bibnamefont {Price}}, \bibinfo
  {author} {\bibfnamefont {E.}~\bibnamefont {Ginossar}}, \bibinfo {author}
  {\bibfnamefont {A.}~\bibnamefont {Nunnenkamp}}, \bibinfo {author}
  {\bibfnamefont {A.~A.}\ \bibnamefont {Houck}}, \bibinfo {author}
  {\bibfnamefont {J.~M.}\ \bibnamefont {Gambetta}}, \bibinfo {author}
  {\bibfnamefont {J.}~\bibnamefont {Koch}}, \bibinfo {author} {\bibfnamefont
  {G.}~\bibnamefont {Johansson}}, \bibinfo {author} {\bibfnamefont {S.~M.}\
  \bibnamefont {Girvin}}, \ and\ \bibinfo {author} {\bibfnamefont {R.~J.}\
  \bibnamefont {Schoelkopf}},\ }\href {\doibase 10.1088/1367-2630/11/7/073040}
  {\bibfield  {journal} {\bibinfo  {journal} {New J. Phys.}\ }\textbf {\bibinfo
  {volume} {11}},\ \bibinfo {pages} {073040} (\bibinfo {year}
  {2009})}\BibitemShut {NoStop}%
\bibitem [{\citenamefont {Helmer}\ and\ \citenamefont
  {Marquardt}(2009)}]{Helmer2009Measurementbased}%
  \BibitemOpen
  \bibfield  {author} {\bibinfo {author} {\bibfnamefont {F.}~\bibnamefont
  {Helmer}}\ and\ \bibinfo {author} {\bibfnamefont {F.}~\bibnamefont
  {Marquardt}},\ }\href {\doibase 10.1103/physreva.79.052328} {\bibfield
  {journal} {\bibinfo  {journal} {Phys. Rev. A}\ }\textbf {\bibinfo {volume}
  {79}},\ \bibinfo {pages} {052328} (\bibinfo {year} {2009})}\BibitemShut
  {NoStop}%
\bibitem [{\citenamefont {Nigg}\ and\ \citenamefont
  {Girvin}(2013)}]{Nigg2013Stabilizer}%
  \BibitemOpen
  \bibfield  {author} {\bibinfo {author} {\bibfnamefont {S.~E.}\ \bibnamefont
  {Nigg}}\ and\ \bibinfo {author} {\bibfnamefont {S.~M.}\ \bibnamefont
  {Girvin}},\ }\href {\doibase 10.1103/physrevlett.110.243604} {\bibfield
  {journal} {\bibinfo  {journal} {Phys. Rev. Lett.}\ }\textbf {\bibinfo
  {volume} {110}},\ \bibinfo {pages} {243604} (\bibinfo {year}
  {2013})}\BibitemShut {NoStop}%
\bibitem [{\citenamefont {Govia}\ \emph {et~al.}(2015)\citenamefont {Govia},
  \citenamefont {Pritchett}, \citenamefont {Plourde}, \citenamefont {Vavilov},
  \citenamefont {McDermott},\ and\ \citenamefont
  {Wilhelm}}]{Govia2015Scalable}%
  \BibitemOpen
  \bibfield  {author} {\bibinfo {author} {\bibfnamefont {L.~C.~G.}\
  \bibnamefont {Govia}}, \bibinfo {author} {\bibfnamefont {E.~J.}\ \bibnamefont
  {Pritchett}}, \bibinfo {author} {\bibfnamefont {B.~L.~T.}\ \bibnamefont
  {Plourde}}, \bibinfo {author} {\bibfnamefont {M.~G.}\ \bibnamefont
  {Vavilov}}, \bibinfo {author} {\bibfnamefont {R.}~\bibnamefont {McDermott}},
  \ and\ \bibinfo {author} {\bibfnamefont {F.~K.}\ \bibnamefont {Wilhelm}},\
  }\href {\doibase 10.1103/physreva.92.022335} {\bibfield  {journal} {\bibinfo
  {journal} {Phys. Rev. A}\ }\textbf {\bibinfo {volume} {92}},\ \bibinfo
  {pages} {022335} (\bibinfo {year} {2015})}\BibitemShut {NoStop}%
\bibitem [{\citenamefont {Viola}\ \emph {et~al.}(1999)\citenamefont {Viola},
  \citenamefont {Knill},\ and\ \citenamefont {Lloyd}}]{Viola1999Dynamical}%
  \BibitemOpen
  \bibfield  {author} {\bibinfo {author} {\bibfnamefont {L.}~\bibnamefont
  {Viola}}, \bibinfo {author} {\bibfnamefont {E.}~\bibnamefont {Knill}}, \ and\
  \bibinfo {author} {\bibfnamefont {S.}~\bibnamefont {Lloyd}},\ }\href
  {\doibase 10.1103/physrevlett.82.2417} {\bibfield  {journal} {\bibinfo
  {journal} {Phys. Rev. Lett.}\ }\textbf {\bibinfo {volume} {82}},\ \bibinfo
  {pages} {2417} (\bibinfo {year} {1999})}\BibitemShut {NoStop}%
\bibitem [{\citenamefont {Paik}\ \emph {et~al.}(2011)\citenamefont {Paik},
  \citenamefont {Schuster}, \citenamefont {Bishop}, \citenamefont {Kirchmair},
  \citenamefont {Catelani}, \citenamefont {Sears}, \citenamefont {Johnson},
  \citenamefont {Reagor}, \citenamefont {Frunzio}, \citenamefont {Glazman},
  \citenamefont {Girvin}, \citenamefont {Devoret},\ and\ \citenamefont
  {Schoelkopf}}]{Paik2011Observation}%
  \BibitemOpen
  \bibfield  {author} {\bibinfo {author} {\bibfnamefont {H.}~\bibnamefont
  {Paik}}, \bibinfo {author} {\bibfnamefont {D.~I.}\ \bibnamefont {Schuster}},
  \bibinfo {author} {\bibfnamefont {L.~S.}\ \bibnamefont {Bishop}}, \bibinfo
  {author} {\bibfnamefont {G.}~\bibnamefont {Kirchmair}}, \bibinfo {author}
  {\bibfnamefont {G.}~\bibnamefont {Catelani}}, \bibinfo {author}
  {\bibfnamefont {A.~P.}\ \bibnamefont {Sears}}, \bibinfo {author}
  {\bibfnamefont {B.~R.}\ \bibnamefont {Johnson}}, \bibinfo {author}
  {\bibfnamefont {M.~J.}\ \bibnamefont {Reagor}}, \bibinfo {author}
  {\bibfnamefont {L.}~\bibnamefont {Frunzio}}, \bibinfo {author} {\bibfnamefont
  {L.~I.}\ \bibnamefont {Glazman}}, \bibinfo {author} {\bibfnamefont {S.~M.}\
  \bibnamefont {Girvin}}, \bibinfo {author} {\bibfnamefont {M.~H.}\
  \bibnamefont {Devoret}}, \ and\ \bibinfo {author} {\bibfnamefont {R.~J.}\
  \bibnamefont {Schoelkopf}},\ }\href {\doibase 10.1103/physrevlett.107.240501}
  {\bibfield  {journal} {\bibinfo  {journal} {Phys. Rev. Lett.}\ }\textbf
  {\bibinfo {volume} {107}},\ \bibinfo {pages} {240501} (\bibinfo {year}
  {2011})}\BibitemShut {NoStop}%
\bibitem [{\citenamefont {Rigetti}\ \emph {et~al.}(2012)\citenamefont
  {Rigetti}, \citenamefont {Gambetta}, \citenamefont {Poletto}, \citenamefont
  {Plourde}, \citenamefont {Chow}, \citenamefont {C\'{o}rcoles}, \citenamefont
  {Smolin}, \citenamefont {Merkel}, \citenamefont {Rozen}, \citenamefont
  {Keefe}, \citenamefont {Rothwell}, \citenamefont {Ketchen},\ and\
  \citenamefont {Steffen}}]{Rigetti2012Superconducting}%
  \BibitemOpen
  \bibfield  {author} {\bibinfo {author} {\bibfnamefont {C.}~\bibnamefont
  {Rigetti}}, \bibinfo {author} {\bibfnamefont {J.~M.}\ \bibnamefont
  {Gambetta}}, \bibinfo {author} {\bibfnamefont {S.}~\bibnamefont {Poletto}},
  \bibinfo {author} {\bibfnamefont {B.~L.~T.}\ \bibnamefont {Plourde}},
  \bibinfo {author} {\bibfnamefont {J.~M.}\ \bibnamefont {Chow}}, \bibinfo
  {author} {\bibfnamefont {A.~D.}\ \bibnamefont {C\'{o}rcoles}}, \bibinfo
  {author} {\bibfnamefont {J.~A.}\ \bibnamefont {Smolin}}, \bibinfo {author}
  {\bibfnamefont {S.~T.}\ \bibnamefont {Merkel}}, \bibinfo {author}
  {\bibfnamefont {J.~R.}\ \bibnamefont {Rozen}}, \bibinfo {author}
  {\bibfnamefont {G.~A.}\ \bibnamefont {Keefe}}, \bibinfo {author}
  {\bibfnamefont {M.~B.}\ \bibnamefont {Rothwell}}, \bibinfo {author}
  {\bibfnamefont {M.~B.}\ \bibnamefont {Ketchen}}, \ and\ \bibinfo {author}
  {\bibfnamefont {M.}~\bibnamefont {Steffen}},\ }\href {\doibase
  10.1103/physrevb.86.100506} {\bibfield  {journal} {\bibinfo  {journal} {Phys.
  Rev. B}\ }\textbf {\bibinfo {volume} {86}},\ \bibinfo {pages} {100506}
  (\bibinfo {year} {2012})}\BibitemShut {NoStop}%
\bibitem [{\citenamefont {Barends}\ \emph {et~al.}(2013)\citenamefont
  {Barends}, \citenamefont {Kelly}, \citenamefont {Megrant}, \citenamefont
  {Sank}, \citenamefont {Jeffrey}, \citenamefont {Chen}, \citenamefont {Yin},
  \citenamefont {Chiaro}, \citenamefont {Mutus}, \citenamefont {Neill},
  \citenamefont {O'Malley}, \citenamefont {Roushan}, \citenamefont {Wenner},
  \citenamefont {White}, \citenamefont {Cleland},\ and\ \citenamefont
  {Martinis}}]{Barends2013Coherent}%
  \BibitemOpen
  \bibfield  {author} {\bibinfo {author} {\bibfnamefont {R.}~\bibnamefont
  {Barends}}, \bibinfo {author} {\bibfnamefont {J.}~\bibnamefont {Kelly}},
  \bibinfo {author} {\bibfnamefont {A.}~\bibnamefont {Megrant}}, \bibinfo
  {author} {\bibfnamefont {D.}~\bibnamefont {Sank}}, \bibinfo {author}
  {\bibfnamefont {E.}~\bibnamefont {Jeffrey}}, \bibinfo {author} {\bibfnamefont
  {Y.}~\bibnamefont {Chen}}, \bibinfo {author} {\bibfnamefont {Y.}~\bibnamefont
  {Yin}}, \bibinfo {author} {\bibfnamefont {B.}~\bibnamefont {Chiaro}},
  \bibinfo {author} {\bibfnamefont {J.}~\bibnamefont {Mutus}}, \bibinfo
  {author} {\bibfnamefont {C.}~\bibnamefont {Neill}}, \bibinfo {author}
  {\bibfnamefont {P.}~\bibnamefont {O'Malley}}, \bibinfo {author}
  {\bibfnamefont {P.}~\bibnamefont {Roushan}}, \bibinfo {author} {\bibfnamefont
  {J.}~\bibnamefont {Wenner}}, \bibinfo {author} {\bibfnamefont {T.~C.}\
  \bibnamefont {White}}, \bibinfo {author} {\bibfnamefont {A.~N.}\ \bibnamefont
  {Cleland}}, \ and\ \bibinfo {author} {\bibfnamefont {J.~M.}\ \bibnamefont
  {Martinis}},\ }\href {\doibase 10.1103/physrevlett.111.080502} {\bibfield
  {journal} {\bibinfo  {journal} {Phys. Rev. Lett.}\ }\textbf {\bibinfo
  {volume} {111}},\ \bibinfo {pages} {080502} (\bibinfo {year}
  {2013})}\BibitemShut {NoStop}%
\bibitem [{\citenamefont {G\"{o}ppl}\ \emph {et~al.}(2008)\citenamefont
  {G\"{o}ppl}, \citenamefont {Fragner}, \citenamefont {Baur}, \citenamefont
  {Bianchetti}, \citenamefont {Filipp}, \citenamefont {Fink}, \citenamefont
  {Leek}, \citenamefont {Puebla}, \citenamefont {Steffen},\ and\ \citenamefont
  {Wallraff}}]{Goppl2008Coplanar}%
  \BibitemOpen
  \bibfield  {author} {\bibinfo {author} {\bibfnamefont {M.}~\bibnamefont
  {G\"{o}ppl}}, \bibinfo {author} {\bibfnamefont {A.}~\bibnamefont {Fragner}},
  \bibinfo {author} {\bibfnamefont {M.}~\bibnamefont {Baur}}, \bibinfo {author}
  {\bibfnamefont {R.}~\bibnamefont {Bianchetti}}, \bibinfo {author}
  {\bibfnamefont {S.}~\bibnamefont {Filipp}}, \bibinfo {author} {\bibfnamefont
  {J.~M.}\ \bibnamefont {Fink}}, \bibinfo {author} {\bibfnamefont {P.~J.}\
  \bibnamefont {Leek}}, \bibinfo {author} {\bibfnamefont {G.}~\bibnamefont
  {Puebla}}, \bibinfo {author} {\bibfnamefont {L.}~\bibnamefont {Steffen}}, \
  and\ \bibinfo {author} {\bibfnamefont {A.}~\bibnamefont {Wallraff}},\ }\href
  {\doibase 10.1063/1.3010859} {\bibfield  {journal} {\bibinfo  {journal} {J.
  Appl. Phys.}\ }\textbf {\bibinfo {volume} {104}},\ \bibinfo {pages} {113904}
  (\bibinfo {year} {2008})}\BibitemShut {NoStop}%
\bibitem [{\citenamefont {Osborn}\ \emph {et~al.}(2007)\citenamefont {Osborn},
  \citenamefont {Strong}, \citenamefont {Sirois},\ and\ \citenamefont
  {Simmonds}}]{Osborn2007FrequencyTunable}%
  \BibitemOpen
  \bibfield  {author} {\bibinfo {author} {\bibfnamefont {K.~D.}\ \bibnamefont
  {Osborn}}, \bibinfo {author} {\bibfnamefont {J.~A.}\ \bibnamefont {Strong}},
  \bibinfo {author} {\bibfnamefont {A.~J.}\ \bibnamefont {Sirois}}, \ and\
  \bibinfo {author} {\bibfnamefont {R.~W.}\ \bibnamefont {Simmonds}},\ }\href
  {\doibase 10.1109/tasc.2007.898544} {\bibfield  {journal} {\bibinfo
  {journal} {IEEE Trans. Appl. Supercond.}\ }\textbf {\bibinfo {volume} {17}},\
  \bibinfo {pages} {166} (\bibinfo {year} {2007})}\BibitemShut {NoStop}%
\bibitem [{\citenamefont {Pierre}\ \emph {et~al.}(2014)\citenamefont {Pierre},
  \citenamefont {Svensson}, \citenamefont {Sathyamoorthy}, \citenamefont
  {Johansson},\ and\ \citenamefont {Delsing}}]{Pierre2014Storage}%
  \BibitemOpen
  \bibfield  {author} {\bibinfo {author} {\bibfnamefont {M.}~\bibnamefont
  {Pierre}}, \bibinfo {author} {\bibfnamefont {I.-M.}\ \bibnamefont
  {Svensson}}, \bibinfo {author} {\bibfnamefont {S.~R.}\ \bibnamefont
  {Sathyamoorthy}}, \bibinfo {author} {\bibfnamefont {G.}~\bibnamefont
  {Johansson}}, \ and\ \bibinfo {author} {\bibfnamefont {P.}~\bibnamefont
  {Delsing}},\ }\href {\doibase 10.1063/1.4882646} {\bibfield  {journal}
  {\bibinfo  {journal} {Appl. Phys. Lett.}\ }\textbf {\bibinfo {volume}
  {104}},\ \bibinfo {pages} {232604} (\bibinfo {year} {2014})}\BibitemShut
  {NoStop}%
\end{thebibliography}%

\end{document}